\documentclass{rspublic}	% loads packages amsmath,amssymb,amsthm
\usepackage{epsfig}		% used for including a PS figure
\makeatletter
\def\@oddhead{{\ }\hfil{\it \@shorttitle}\hfil\rm \thepage}
\makeatother
\usepackage[sectionbib]{natbib}

\bibpunct[, ]{(}{)}{;}{a}{}{,}

\makeatletter
\renewcommand\NAT@bibsetup%
   [1]{\setlength\labelwidth{0pt}\setlength\leftmargin{1em}%
       \setlength{\itemindent}{-1em}\setlength{\itemsep}{-3pt}}
\makeatother

\let\mycite\citep
\let\myciteasnoun\citet

\newcommand{\defeq}{:=}

\theoremstyle{plain}
\newtheorem{conjecture}[theorem]{Conjecture}
\newtheorem*{theoremL}{Theorem~L\null}

\theoremstyle{definition}
\newtheorem*{definition}{Definition}

\theoremstyle{remark}
\newtheorem*{remark}{Remark}

\begin{document}

\title[Lam\'e polynomials, hyperelliptic reductions and Lam\'e band structure]{Lam\'e polynomials, hyperelliptic reductions and Lam\'e band structure}
\author[R. S. Maier]{Robert S. Maier} 
\affiliation{Depts.\ of Mathematics and Physics, University of Arizona, Tucson AZ 85721, USA}
\label{firstpage}
\maketitle

% abstract, including (since this is an instance of the rspublic document class)
% a list of keywords as an argument
\begin{abstract}{Lam\'e equation, Lam\'e polynomial, dispersion relation, band structure, hyperelliptic reduction,
Hermite--Krichever Ansatz} The band structure of the Lam\'e equation,
viewed as a one-dimensional Schr\"odinger equation with a periodic
potential, is studied.  At~integer values of the degree parameter~$\ell$,
the dispersion relation is reduced to the $\ell=1$ dispersion relation, and
a previously published $\ell=2$ dispersion relation is shown to be partly
incorrect.  The Hermite--Krichever Ansatz, which expresses Lam\'e equation
solutions in~terms of ${\ell=1}$ solutions, is the chief tool.  It~is based
on a projection from a genus-$\ell$ hyperelliptic curve, which parametrizes
solutions, to an elliptic curve.  A~general formula for this covering is
derived, and is used to reduce certain hyperelliptic integrals to elliptic
ones.  Degeneracies between band edges, which can occur if the Lam\'e
equation parameters take complex values, are investigated.  If the Lam\'e
equation is viewed as a differential equation on an elliptic curve, a
formula is conjectured for the number of points in elliptic moduli space
(elliptic curve parameter space) at which degeneracies occur.  Tables of
spectral polynomials and Lam\'e polynomials, i.e., band edge solutions, are
given.  A~table in the older literature is corrected.
\end{abstract}

% PACS and AMS numbers, if needed
%\pacs{02.30.Gp, 02.30.Ik, 71.20.-b}
%\ams{33E10, 34L40, 14K25}

\section{Introduction}
\label{sec:intro}
\subsection{Background}

The term `Lam\'e equation' refers to any of several closely related
second-order ordinary differential equations
\mycite{Whittaker27,Erdelyi53,Arscott64}.  The first such equation was
obtained by Lam\'e, by applying the method of separation of variables to
Laplace's equation in an ellipsoidal coordinate system.  Lam\'e equations
arise elsewhere in theoretical physics.  Recent application areas include
(i)~the analysis of preheating after inflation, arising from parametric
amplification \mycite{Boyanovsky96,Greene97,Kaiser98,Ivanov2001}; (ii)~the
stability analysis of critical droplets in bounded spatial domains
\mycite{Maier02}; (iii)~the stability analysis of static configurations in
Josephson junctions \mycite{Caputo2000}, (iv)~the computation of the
distance--redshift relation in inhomogeneous cosmologies
\mycite{Kantowski2001}, and (v)~magnetostatic problems in triaxial ellipsoids
\mycite{Dobner98}.

In some versions of the Lam\'e equation, elliptic functions appear
explicitly, and in others (the algebraic versions) they appear implicitly.
The version that appears most often in the physics literature is the Jacobi
one,
\begin{equation}
\label{eq:jacobilame}
\left[ -\frac{\rd^2}{\rd\alpha^2} + \ell(\ell+1)m\,{\rm sn}^2(\alpha|m)
\right] \Psi = E\,\Psi,
\end{equation}
which is a one-dimensional Schr\"odinger equation with a doubly periodic
potential, parametrized by $m$ and~$\ell$.  Here ${\rm sn}(\cdot|m)$ is the
Jacobi elliptic function with modular parameter~$m$.  $m$~is often
restricted to~$(0,1)$, though in~general $m\in\mathbb{C}\setminus\{0,1\}$
is allowed.  When $m\in(0,1)$, the function ${\rm sn}^2(\cdot|m)$ has real
period $2K\defeq2{\mathsf{K}}(m)$ and imaginary period $2\ri
K'\defeq2\ri{\sf K}(1-m)$, with ${\mathsf{K}}(m)$ the first complete
elliptic integral.

If $\alpha$ is restricted to the real axis and $m$~and~$\ell$ are real,
(\ref{eq:jacobilame})~becomes a real-domain Schr\"odinger equation with a
periodic potential, i.e., a Hill's equation.  Standard results on Hill's
equation apply \mycite{Magnus79,McKean79}.  Equipping~(\ref{eq:jacobilame})
with a quasi-periodic boundary condition
\begin{equation}
\Psi(\alpha+2K) = \re^{\ri k(2K)}\Psi(\alpha) \defeq \xi\,\Psi(\alpha),
\end{equation}
where $k\in\mathbb{R}$ is fixed, defines a self-adjoint boundary value
problem.  For any real~$k$ (i.e., for any Floquet multiplier~$\xi$ with
$|\xi|=1$), there will be an infinite discrete set of
energies~$E\in\mathbb{R}$ for which this problem has a solution, called a
Bloch solution with crystal momentum~$k$.  Each such~$E$ will lie in one of
the allowed zones, which are intervals delimited by energies corresponding
to $\xi=\pm1$, i.e., to periodic and anti-periodic Bloch solutions.  These
form a sequence $E_0 < E_1 \le E_2 < E_3 \le E_4 < \cdots$, where $E_0$~is
a `periodic' eigenvalue, followed by alternating pairs of anti-periodic and
periodic eigenvalues (each pair may be coincident).  The allowed zones are
the intervals $[E_{2j},E_{2j+1}]$.  The complementary intervals
$(E_{2j+1},E_{2j+2})$ are forbidden zones, or lacun\ae.  Any solution of
Hill's equation with energy in a lacuna is unstable: its multiplier~$\xi$
will not satisfy $|\xi|=1$, and its crystal momentum~$k$ will not be real.

It is a celebrated result of~\myciteasnoun{Ince40b} that if the degree $\ell$
is an integer, which without loss of generality may be chosen to be
non-negative, the Lam\'e equation~(\ref{eq:jacobilame}) will have only a
finite number of nonempty lacun\ae.  A~converse to this statement holds
as~well \mycite{Gesztesy95}.  If $\ell$~is an integer, the Bloch spectrum
consists of the $\ell+1$ bands
$[E_0,E_1],[E_2,E_3],\dots,[E_{2\ell},\infty)$, and $\ell(\ell+1)m\,{\rm
sn}(\cdot|m)$ is said to be a {\em finite-band\/} or algebro-geometric
potential.  The $2\ell+1$ band edges $E_0,\dots,E_{2\ell}$ are algebraic
functions of the parameter~$m$.  That~is, they are the roots of a certain
polynomial, the coefficients of which are polynomial in~$m$.  The
corresponding periodic and anti-periodic Bloch solutions are called {\em
Lam\'e polynomials\/}: they are polynomials in the Jacobi elliptic
functions ${\rm sn}(\alpha|m)$, ${\rm cn}(\alpha|m)$, and ${\rm
dn}(\alpha|m)$.  The double eigenvalues embedded in the topmost band
$[E_{2\ell},\infty)$ (the `conduction' band), namely $E_{2j}=E_{2j+1}$,
$j>\ell$, are loosely called transcendental eigenvalues.  For $\ell=1$
at~least, they are known to be transcendental functions of~$m$
\mycite{Chudnovsky80}.

There has been much work on {\em algebraizing\/} the integer-$\ell$ Lam\'e
equation, to facilitate the computation of the band edges and the
coefficients of the Lam\'e polynomials 
\mycite{Alhassid83,Turbiner89,Li99,Li00,Finkel2000}.  Such schemes have been
extended to the case when $\ell$~is a half-odd-integer, in which there are
an infinite number of lacun\ae.  In this case, certain `mid-band' Bloch
functions, namely ones with $\xi=\pm\ri$ and real period~$8K$, are
algebraic functions of ${\rm sn}(\alpha|m)$.  Certain rational values
of~$\ell$ with $2\ell\notin\mathbb{Z}$ also yield algebraic Bloch
functions, provided the parameters $m$ and~$E$ are chosen appropriately
\mycite{Maier04}.

An algebraic understanding of band edges is useful, but it is also
desirable to have a closed-form expression for the {\em dispersion
relation\/}: $k$~as a function of~$E$.  The value of~$k$ is not unique,
since it can be negated (equivalently, $\xi\mapsto1/\xi$), and any integer
multiple of~$\pi/K$ can be added.  However, each branch has the property
that $k\sim E^{1/2}$ or $k\sim- E^{1/2}$ to leading order as $E\to+\infty$.
Also, $k\in\mathbb{R}$ in each band.

The goal of this paper is the efficient computation of the dispersion
relation when $\ell$~is an integer.  The following are illustrations of why
this is of importance in theoretical physics.  In application~(i) above
(preheating after inflation), particle production is due to parametric
amplification: a~solution having a multiplier~$\xi$ with $|\xi|>1$.  This
corresponds to the energy~$E$ not being at a band edge, or even in a band,
but in a lacuna.  In application~(ii) (the stability analysis of a critical
droplet), the analysis includes an imposition of Dirichlet rather than
quasi-periodic boundary conditions on an $\ell=2$ Lam\'e equation
\mycite{Maier02}.  The resulting Bloch solution is not a Lam\'e polynomial,
but rather a mid-band solution.

When $\ell$~is an integer, the Lam\'e equation is integrable, and the
general integral of~(\ref{eq:jacobilame}) can be expressed in~terms of
Jacobi theta functions.  The dispersion relations $k=k_\ell(E|m)$,
$\ell\ge1$, can in~principle be computed in~terms of elliptic integrals.
The case $\ell=1$ is by far the easiest.  If $\ell=1$, the solution space
of~(\ref{eq:jacobilame}), except when $E$~is at a band edge, will be
spanned by the pair of Hermite--Halphen solutions
\begin{equation}
\label{eq:jacobiPhi}
\widetilde\Phi(\alpha;\pm\alpha_0|m) \defeq \frac{{\rm
H}(\alpha\mp\alpha_0|m)}{\Theta(\alpha|m)}\exp\left[\alpha\,{\rm
Z}(\pm\alpha_0|m)\right].
\end{equation}
Here ${\rm H},\Theta,{\rm Z}$ are the Jacobi eta, theta and zeta functions,
with periods $4K,2K,2K$ respectively, and $\alpha_0\in\mathbb{C}$ is
defined up~to sign by ${\rm dn}^2(\alpha_0|m)=E-m$.  So
\begin{equation}
k_1(E|m)=-\ri\,{\rm Z}(\alpha_0|m)+\pi/2\mathsf{K}(m),\qquad {\rm dn}^2(\alpha_0|m)=E-m,
\end{equation}
up~to multi-valuedness.  This is a {\em parametric\/} dispersion relation.
It has been exploited in a study of Wannier--Stark resonances with
non-real~$E$ and~$k$ \mycite{Grecchi97,Sacchetti97}.  However, the extension
to $\ell>1$ is numerically nontrivial.  $k_\ell(E|m)$ turns out to equal
$\sum_{r=0}^{\ell-1} [-\ri\,{\rm Z}(\alpha_r|m)+\pi/2\mathsf{K}(m)]$, where
$\{\alpha_r\}_{r=0}^{\ell-1}$ satisfy coupled transcendental equations
involving $E$~and~$m$ \mycite[\S\,23.71]{Whittaker27}.  Li, Kusnezov \&
Iachello calculated and graphed $k=k_2(E|m)$ as~well as $k=k_1(E|m)$ in the
`lemniscatic' case $m=\frac12$ \mycite{Li99,Li00}.  Unfortunately their graph
of $k=k_2(E|\frac12)$ is incorrect, as will be shown.

\subsection{Overview of results}

When $\ell>1$, we abandon the traditional Hermite--Halphen solutions, and
examine instead the implications for the Lam\'e dispersion relations of
what is now called the Hermite--Krichever Ansatz.  This is an alternative
way of generating closed-form solutions of the Lam\'e solution at arbitrary
energy~$E$; for small integer values of~$\ell$, at~least.  Until the 1980s,
the only reference for the Ansatz was the classic work of
\myciteasnoun[chapitre~XII]{Halphen1886}, who applied it to the cases
$\ell=2,3,4$, and in~part to $\ell=5$.  \myciteasnoun{Krichever80orig}
revived it as an aid in the construction of elliptic solutions of the
Korteweg--de~Vries and other integrable evolution equations.
\myciteasnoun{Belokolos86orig} and \myciteasnoun{Belokolos2002orig} summarize
early and recent developments.

The Hermite--Krichever Ansatz is easy to explain, even in the context of
the Jacobi form of the Lam\'e equation, which is not the most convenient
for symbolic manipulation.  It~asserts that for any integer $\ell\ge1$,
fundamental solutions of~(\ref{eq:jacobilame}) can be constructed from the
$\ell=1$ solutions $\widetilde\Phi(\alpha;\pm\alpha_0|m)$ as finite series
of the form
\begin{equation}
\label{eq:jacobiHK}
\left[\sum_{j=0}^{N_\ell} C_j^{(\ell)}
\frac{\rd^j}{\rd\alpha^j} \widetilde\Phi(\alpha;\pm\alpha_0|m)\right]
\exp(\pm\kappa_\ell\alpha),
\end{equation}
where the parameter~$\alpha_0$ is now computed from a {\em reduced
energy\/} ${\cal E}_\ell$ by the formula ${\rm dn}^2(\alpha_0|m) = {\cal
E}_\ell-m$.  The reduced energy~${\cal E}_\ell$, the
exponent~$\kappa_\ell$, and the coefficients $\{C_j^{(\ell)}\}$ will depend
on $E$ and~$m$.  ${\cal E}_\ell(E|m)$~may be chosen to be rational in $E$
and~$m$, and $\pm\kappa_\ell$~to be of the form $\hat\kappa_\ell(E|m)$
times $\pm\ri\sqrt{\widetilde{\mathsf{L}}_\ell(E|m)}$, where
$\hat\kappa_\ell(E|m)$ is also rational, and
$\widetilde{\mathsf{L}}_\ell(E|m)$ is the spectral polynomial
$\prod_{s=0}^{2\ell}[E-E_s(m)]$, a~degree-$(2\ell+1)$ polynomial in~$E$ the
coefficients of which, as noted, are rational in~$m$.

If the Lam\'e equation can be integrated in the framework of the
Hermite--Krichever Ansatz, it follows from~(\ref{eq:jacobiHK}) that up~to
sign, etc.,
\begin{equation}
\label{eq:twoterms}
k_\ell(E|m)=k_1({\cal E}_\ell(E|m)|m)+ \hat\kappa_\ell(E|m)
\sqrt{\widetilde{\mathsf{L}}_\ell(E|m)}.
\end{equation}
{\em The dispersion relation for any integer~$\ell$ can be expressed
in~terms of the $\ell=1$ relation.}  To~compute~$k_\ell$, only one
transcendental function (i.e.,~$k_1$) needs to be evaluated, since the
other functions in~(\ref{eq:twoterms}) are elementary.  The only difficult
matter is choosing the relative sign of the two terms, since each is
defined only up~to sign.

The functions ${\cal E}_\ell(E|m),\hat\kappa_\ell(E|m)$ are rational with
integer coefficients, but working them out when $\ell$~is large is a
lengthy task.  In~principle one can write down a recurrence relation for
the coefficients~$\{C_j\}$, and work~out ${\cal
E}_\ell(E|m),\allowbreak\hat\kappa_\ell(E|m)$ from the condition that the
series terminate.  However, their numerator and denominator degrees grow
quadratically as $\ell$~increases.  This explains why Halphen's treatment
of the $\ell=5$ case was only partial.  In~a series of papers, Kostov,
Enol'skii, and collaborators used computer algebra systems to perform a
full analysis of the cases $\ell=2,3,4,5$
\mycite{Gerdt89,Kostov93orig,Enolskii94,Eilbeck94}.  When $\ell=5$, using
{\sc Mathematica} to compute the integer coefficients of rational functions
equivalent to ${\cal E}_\ell(E|m),\hat\kappa_\ell(E|m)$ required seven
hours of time on a Sparc-1, a~Unix workstation of that era
\mycite{Eilbeck94}.  Until now, their analysis has not been extended to
higher~$\ell$.

While performing extensive symbolic computations, we recently made a
discovery which is formalized in Theorem~L below\null.  {\em For all
integer $\ell\ge2$, the degree-$\ell$ Lam\'e equation can be integrated in
the framework of the Hermite--Krichever Ansatz, and the rational functions
that perform the reduction to the $\ell=1$ case can be computed by simple
formulas from certain spectral polynomials of the degree-$\ell$ equation,
which are relatively easy to work~out.}  These are the ordinary spectral
polynomial $\prod_{s=0}^{2\ell}[E-E_s(m)]$ associated with the band-edge
solutions, and the spectral polynomials associated with two other types of
closed-form solution that have not previously been studied in the
literature.  We~call them {\em twisted\/} and {\em theta-twisted\/} Lam\'e
polynomials.  In~the context of the Jacobi form, the former are polynomials
in ${\rm sn}(\alpha|m)$, ${\rm cn}(\alpha|m)$ and ${\rm dn}(\alpha|m)$,
multiplied by a factor $\exp(\kappa\,\alpha)$.  (If $\kappa\in\mathbb{R}$,
`canted' would be better than `twisted'.)  The latter contain a factor
resembling~(\ref{eq:jacobiPhi}).

Theorem~L follows from modern finite-band integration theory: specifically,
from the Baker--Akhiezer uniformization of the relation between the energy
and the crystal momentum.  This uniformization is closely tied to classical
work on the Lam\'e equation (the parametrized Baker--Akhiezer solutions of
the integer-$\ell$ Lam\'e equation are in~fact equivalent to the
Hermite--Halphen solutions).  Theorem~L greatly simplifies the computation
of higher-$\ell$ dispersion relations.  It~also has implications for the
theory of {\em hyperelliptic reduction\/}: the reduction of hyperelliptic
integrals to elliptic ones \mycite{Belokolos86orig}.  This is on~account of
the following.  For any $\ell\ge1$ and~$m$, the solutions of the Lam\'e
equation, both the Hermite--Halphen solutions and those derived from the
Hermite--Krichever Ansatz, are single-valued functions not of~$E$, but
rather of a point $(E,\tilde\nu)$ on the $\ell$'th Lam\'e {\em spectral
curve\/}: a~hyperelliptic curve comprising all $(E,\tilde\nu)$ satisfying
\begin{equation}
\tilde\nu^2 = \prod_{s=0}^{2\ell}\left[E-E_s(m)\right].
\end{equation}
The fact that $\tilde\nu=\tilde\nu(E)$ is two-valued (except at a band
edge) is responsible for the two-valuedness of, e.g., the parameter
$\alpha_0=\alpha_0(E)$ of~(\ref{eq:jacobiPhi}), and in general, for the
uncertainty in the sign of~$k$.  The $\ell$'th spectral curve generically
has genus~$\ell$ and may be
denoted~$\widetilde\Gamma_\ell\defeq\widetilde\Gamma_\ell(m)$.  For any
integer $\ell\ge2$, the $E\mapsto {\cal E}_\ell(E|m)$ reduction map of the
Hermite--Krichever Ansatz induces a covering
$\pi_\ell:\widetilde\Gamma_\ell\to\widetilde\Gamma_1$.  The first known
covering of an elliptic curve by a higher-genus hyperelliptic curve was
constructed by Legendre and generalized by Jacobi
\mycite[\S\,2]{Belokolos86orig}.  But it is difficult to enumerate such
coverings, or even work~out explicit examples.  Those generated by the
Ansatz applied to the Lam\'e equation are a welcome exception.

The integral with respect to~$E$ of any rational function of $E$
and~$\sqrt{\Pi(E)}$, where $\Pi$~is a polynomial, is a line integral on the
algebraic curve defined by $\tilde\nu^2=\Pi(E)$.  The covering
$\pi_\ell:\widetilde\Gamma_{\ell}\to\widetilde\Gamma_1$, a formula for
which is provided by Theorem~L, reduces certain such hyperelliptic
integrals to elliptic ones.  In modern language, the theorem specifies how
certain holomorphic differentials on hyperelliptic curves can arise as
pullbacks of holomorphic differentials on elliptic curves.

We investigate the degeneracies of the band edges
$\{E_s(m)\}_{s=0}^{2\ell}$ that can occur when the modular parameter~$m$ is
non-real.  Such level-crossings were first considered
by~\myciteasnoun{Cohn1888} in a dissertation that seems not to have been
followed up, though it was later cited by
\myciteasnoun[\S\,23.41]{Whittaker27}.  We~conjecture a formula for the
$\ell$-dependence of the number of values of
$m\in\mathbb{C}\setminus\{0,1\}$, or equivalently the number of values of
the Klein invariant~$J\in\mathbb{C}$, at~which two band edges coincide.
When this occurs, the genus of the hyperelliptic spectral
curve~$\widetilde\Gamma_\ell$ is reduced from $\ell$ to~$\ell-1$, though
its {\em arithmetic\/} genus remains equal to~$\ell$.  Band-edge
degeneracies are responsible for a fact discovered by
\myciteasnoun{Turbiner89}: if~$\ell\ge2$, the complex curve comprising all
points $\{(m,E_s(m))\}_{s=0}^{2\ell}$
in~$\mathbb{C}\setminus\{0,1\}\times\mathbb{C}$ has only four, rather than
$2\ell+1$, connected components.

This paper is organized as follows.  Section~\ref{sec:algebraic}~introduces
the Lam\'e equation in its elliptic-curve form and relates the
Hermite--Halphen solutions to the Baker--Akhiezer function.
In~\S\,\ref{sec:families}, Lam\'e polynomials in the context of the
elliptic-curve form are classified.  In~\S\,\ref{sec:HK}, the
Hermite--Krichever Ansatz is introduced, and Theorem~L is stated and
proved.  The application to hyperelliptic reduction is covered
in~\S\,\ref{sec:reduction}.  In \S\S\,\ref{sec:dispersion}
and~\ref{sec:jacobi}, dispersion relations are worked~out and the
abovementioned dispersion relation for the case $\ell=2$ is corrected.  The
$\ell=3$ dispersion relation is graphed as~well.  Finally, an area for
future investigation is mentioned in~\S\,\ref{sec:conclusions}.

\section{The elliptic-curve algebraic form}
\label{sec:algebraic}

In \S\S\,\ref{sec:families} through~\ref{sec:dispersion} we use exclusively
what we call the elliptic-curve algebraic form of the Lam\'e equation,
which is the most convenient for symbolic computation.  In~this section we
derive~it, and also define a fundamental multi-valued function~$\Phi$,
which appears in the elliptic-curve version of both the Hermite--Halphen
solutions and the Hermite--Krichever Ansatz.

\subsection{An elliptic-curve Schr\"odinger equation}

Many algebraic forms can be obtained from~(\ref{eq:jacobilame}) by changing
to new independent variables which are elliptic functions of~$\alpha$, such
as ${\rm sn}(\alpha|m)$.  (See, e.g., \myciteasnoun[\S\,1.1]{Arscott62} and
\myciteasnoun[pp~192--3]{Arscott64}).  A~form in which the domain of
definition is explicitly a cubic algebraic curve of genus~$1$, i.e., a
cubic elliptic curve, can be obtained as follows.  First, the Lam\'e
equation is restated in terms of the Weierstrassian
function~$\wp=\wp(u;g_2,g_3)$.  This is the canonical elliptic function
with a double pole at~$u=0$, satisfying $(\wp')^2=f(\wp)$ where
$f(x)\defeq4x^3-g_2x-g_3=4\prod_{\gamma=1}^3(x-e_\gamma)$.  For ellipticity
the roots $\{e_\gamma\}_{\gamma=1}^3$ must be distinct, which is equivalent
to the condition that the modular discriminant $\Delta\defeq g_2^3-27g_3^2$
be non-zero.  Either of $g_2,g_3\in\mathbb{C}$ may equal zero, but not
both.

The relation between the Jacobi and Weierstrass elliptic functions is
well-known \mycite[\S\,18.9]{Abramowitz65}.  Choose $\{e_\gamma\}_{\gamma=1}^3$
according to
\begin{equation}
\label{eq:e1e2e3}
(e_1,e_2,e_3)=A^2\left(\frac{2-m}3,\frac{2m-1}3,\frac{-(m+1)}3\right),
\end{equation}
where $A\in\mathbb{C}\setminus\{0\}$
is any convenient proportionality constant.  Then
\begin{equation}
\label{eq:g2g3}
g_2=A^4\frac{4(m^2-m+1)}3,\qquad
g_3=A^6\frac{4(m-2)(2m-1)(m+1)}{27},
\end{equation}
and the dimensionless ($A$-independent) Klein invariant $J\defeq g_2^3/\Delta$
will be given by
\begin{equation}
\label{eq:mtoJ}
J=\frac{4}{27}\frac{(m^2-m+1)^3}{m^2(1-m)^2}.
\end{equation}
The two sorts of elliptic function will be related by, e.g.,
\begin{equation}
\label{eq:snns}
{\rm sn}^2(Az|m)=\frac{e_1-e_3}{\wp(z)-e_3},\qquad
{\rm ns}^2(Az|m)=\frac{\wp(z)-e_3}{e_1-e_3},
\end{equation}
and the periods of~$\wp$, denoted $2\omega,2\omega'$, will be related to
those of ${\rm sn}^2$ by
\begin{equation}
2\omega=2K/A,\qquad 2\omega'=2\ri K'/A.
\end{equation}
The case when $2K,2K'$ are real, or equivalently $\omega\in\mathbb{R}$,
$\omega'\in \ri\mathbb{R}$ (we~assume $A\in\mathbb{R}$), is the case when
$g_2,g_3\in\mathbb{R}$ and $\Delta>0$ \mycite[\S\,18.1]{Abramowitz65}.

Choosing for simplicity $A=1$, so that $e_1-e_3=A^2=1$, and rewriting the
Lam\'e equation~(\ref{eq:jacobilame}) with the aid of~(\ref{eq:snns}),
yields the {\em Weierstrassian form\/}
\begin{equation}
\label{eq:weierstrasslame}
\left\{\frac{\rd^2}{\rd u^2} - \left[\ell(\ell+1)\wp(u;g_2,g_3) + B\right]\right\}\Psi=0,
\end{equation}
where $u\defeq\alpha+\ri K'$.  (The translation of~(\ref{eq:jacobilame})
by~$\ri K'$ replaces $m\,{\rm sn}^2$ by~${\rm ns}^2$.)  Here
$B\defeq-E(e_1-e_3)-\ell(\ell+1)e_3$, i.e.,
\begin{equation}
\label{eq:BE}
B\defeq-E+\tfrac13\ell(\ell+1)(m+1),
\end{equation}
is a transformed energy parameter.  Changing to the new independent
variable $x\defeq\wp(u;g_2,g_3)$ converts (\ref{eq:weierstrasslame}) to the
commonly encountered algebraic form
\begin{equation}
\label{eq:alglame}
\left\{
\frac{\rd^2}{\rd x^2} + \frac12 \sum_{\gamma=1}^3
\frac1{x-e_\gamma}\frac{\rd}{\rd x} - \frac {\ell(\ell+1)x +
B}{4\prod_{\gamma=1}^3(x-e_\gamma)}
\right\}\Psi=0.
\end{equation}
This is a differential equation on the Riemann sphere
$\mathbb{P}^1\defeq\mathbb{C}\cup\{\infty\}$ with regular singular points
at $x=e_1,e_2,e_3,\infty$.  Any solution of the original Lam\'e
equation~(\ref{eq:jacobilame}) or the Weierstrassian
form~(\ref{eq:weierstrasslame}) which is quasi-periodic in the sense that
it is multiplied by $\xi,\xi'\in\mathbb{C}\setminus\{0\}$ when
$\alpha\leftarrow\alpha+2K$ or $\alpha\leftarrow\alpha+2\ri K'$
respectively (equivalently, when $u\leftarrow u+2\omega$ or $u\leftarrow
u+2\omega'$), will be a {\em path-multiplicative\/} function of~$x$.
That~is, it will be multiplied by~$\xi,\xi'$ when it is analytically
continued around a cut joining the pair of points $x=e_2,e_3$ or
$x=e_1,e_2$, respectively.  In~the context of the algebraic form, the
dispersion relation is still a relation between the energy and a
multiplier~$\xi$, but the multiplier is interpreted as specifying not
quasi-periodicity on~$\mathbb{C}$, but rather multi-valuedness
on~$\mathbb{P}^1$.

The algebraic form~(\ref{eq:alglame}) of the Lam\'e equation lifts
naturally to the complex elliptic curve
$E_{g_2,g_3}\defeq\{(x,y)\in\mathbb{C}^2\mid
y^2=f(x)\}\cup\{(\infty,\infty)\}$ over~$\mathbb{P}^1$, parametrized by
$(x,y)\defeq\left(\wp(u),\wp'(u)\right)$.  One may rewrite
(\ref{eq:alglame}) in the {\em elliptic-curve algebraic form\/}
\begin{equation}
\label{eq:ellipticlame}
\left\{
\left(y\frac{\rd}{\rd x}\right)^2 - \left[\ell(\ell+1)x + B\right]
\right\}\Psi=0.
\end{equation}
This is an elliptic-curve Schr\"odinger equation of the form
\begin{equation}
\label{eq:ellipticSchroedinger}
\left[-
\left(y\frac{\rd}{\rd x}\right)^2 + q(x)
\right]\Psi= -B\,\Psi,
\end{equation}
with the (rational) potential function $q(x)$ taken to equal
$\ell(\ell+1)x$.  Equation~(\ref{eq:ellipticlame}) follows directly
from~(\ref{eq:weierstrasslame}), since $\rd/\rd u=\wp'\,\rd/\rd\wp=
y\,\rd/\rd x$.  It~is a differential equation on~$E_{g_2,g_3}$ with a
single singular point: a~regular one at $(x,y)=(\infty,\infty)$.  Note that
the $2$-to-$1$ covering map $\pi:E_{g_2,g_3}\to\mathbb{P}^1$ defined by
$\pi(x,y)=x$ has $\{(e_\gamma,0)\}_{\gamma=1}^3$ and $(\infty,\infty)$ as
simple critical points.  One reason why (\ref{eq:ellipticlame}) is more
fundamental than~(\ref{eq:alglame}) is that the singular points
of~(\ref{eq:alglame}) at $x=e_1,e_2,e_3$ can be regarded as artifacts:
consequences of $\{(e_\gamma,0)\}_{\gamma=1}^3$ being critical points
of~$\pi$.

The complex-analytic differential geometry of the elliptic curve
$E_{g_2,g_3}$ takes a bit of getting used~to.  Both $x$ and~$y$ are
meromorphic $\mathbb{P}^1$-valued functions on~$E_{g_2,g_3}$, and the only
pole that either has on~$E_{g_2,g_3}$ is at the point
$O\defeq(\infty,\infty)$.  In~a neighborhood of any generic point $(x,y)$
other than $O$ and the three points $(e_\gamma,0)$, either $x$ or~$y$ will
serve as a local coordinate.  However, near each~$(e_\gamma,0)$ only~$y$
will be a good local coordinate, since $\rd y/\rd x$ diverges at
$x=e_\gamma$.  Also, $x$~has a double and $y$~has a triple pole at~$O$, so
the appropriate local coordinate near~$O$ is the quotient~$x/y$.  The
$1$-form $\rd x/y$ is not merely meromorphic but {\em holomorphic\/}, with
no poles on~$E_{g_2,g_3}$.  Its dual is the vector field (or~directional
derivative) $y\,\rd/\rd x$.

Elliptic functions, i.e., doubly periodic functions, of the original
variable $u\in\mathbb{C}$ correspond to single-valued functions
on~$E_{g_2,g_3}$.  These are rational functions of $x,y$ and may be written
as $R_0(x)+R_1(x)y$, i.e., $R_0(\wp(u))+R_1(\wp(u))\wp'(u)$.  The formula
$(y\,\rd/\rd x)y=6x^2-\frac12 g_2$ allows such functions to be
differentiated algebraically.  In~a similar way, quasi-doubly periodic
functions of~$u$ (sometimes called {\em elliptic functions of the second
kind\/}), which are multiplied by~$\xi$ when $u\leftarrow u+2\omega$ and
by~$\xi'$ when $u\leftarrow u+2\omega'$, correspond to multiplicatively
multi-valued functions on~$E_{g_2,g_3}$.

$E_{g_2,g_3}$ has genus~$1$ and is topologically a torus.  A~fundamental
pair of loops that cannot be shrunk to a point may be chosen to be a loop
that extends between $(e_2,0)$ and~$(e_3,0)$, and one that extends between
$(e_1,0)$ and~$(e_2,0)$, with (if~$g_2,g_3$ are real, at~least) half of
each loop passing through positive values of~$y$, and the other half
through negative values.  One way of constructing an elliptic function of
the second kind is to anti-differentiate a rational function $R(x,y)$.
Here `anti-differentiate' means to compute $\int R(x,y)\,\rd x/y$, its
indefinite integral against the holomorphic $1$-form~$\rd x/y$.  The
resulting function will typically have a non-zero modulus of periodicity
associated with each loop.  If~so, exponentiating~it will yield a
multiplicatively multi-valued function on~$E_{g_2,g_3}$, with non-unit
multipliers~$\xi,\xi'$.

\subsection{The Hermite--Halphen solutions}

The fundamental multi-valued function~$\Phi$ on the elliptic curve will now
be defined.  It~is an elliptic-curve version of Halphen's {\em
l'\'el\'ement simple\/} \mycite{Halphen1886}.

\begin{definition}
On the elliptic curve $E_{g_2,g_3}$, the multi-valued meromorphic
function~$\Phi$, parametrized by $(x_0,y_0)\in
E_{g_2,g_3}\setminus\{(\infty,\infty)\}$, is defined up~to a constant
factor by a formula containing an indefinite elliptic integral,
\begin{equation}
\label{eq:Phi}
\Phi(x,y;x_0,y_0) = 
\exp\left[\frac12 \int \left(\frac{y+y_0}{x-x_0}\right)\frac{\rd x}y\right].
\end{equation}
Its multi-valuedness, which is multiplicative, arises from the path of
integration winding around $E_{g_2,g_3}$ in any combination of the two
directions.  Each branch of~$\Phi$~has a simple zero at $(x,y)=(x_0,y_0)$
and a simple pole at $(x,y)=(\infty,\infty)$.
\end{definition}

To motivate the definition of~$\Phi$, a brief sketch will now be given of
the construction of the Hermite--Halphen solutions of the elliptic-curve
algebraic Lam\'e equation~(\ref{eq:ellipticlame}) for integer $\ell\ge1$.
The standard published exposition is not fully algebraic, being framed
largely in the context of the Weierstrassian form
\mycite[\S\,23.7]{Whittaker27}.  This sketch will relate the Hermite--Halphen
solutions to modern finite-band integration theory and the concept of a
Baker--Akhiezer function \mycite{Gesztesy2003,Treibich2001orig}.  The
starting point is the differential equation
\begin{equation}
\label{eq:pseudohalphen}
\left\{
\left(y\frac{\rd}{\rd x}\right)^3 - 4[q(x)+B]\left(y\frac{\rd}{\rd
x}\right) - 2\left[\left(y\frac{\rd}{\rd x}\right)q(x)\right]
\right\}\mathcal{F} = 0.
\end{equation}
The differential operator in~(\ref{eq:pseudohalphen}) is the `symmetric
square' of the elliptic-curve Schr\"odinger operator
of~(\ref{eq:ellipticSchroedinger}), so the solutions
of~(\ref{eq:pseudohalphen}) include the product of any pair of solutions
of~(\ref{eq:ellipticSchroedinger}).  If~the potential function~$q(x)$ is
rational, it is known that the solution space of~(\ref{eq:pseudohalphen})
contains a function~${\cal F}(x;B)$ which is (i)~meromorphic in~$x$ (the
only poles being at the poles of~$q(x)$) and (ii)~monic polynomial in~$B$,
if~and only~if (\ref{eq:ellipticSchroedinger})~is a finite-band
Schr\"odinger equation on~$E_{g_2,g_3}$, i.e., a finite-band Schr\"odinger
equation with a doubly periodic potential~\mycite{Its74orig}.  This is an
alternative to the characterization of \myciteasnoun{Gesztesy96}, according
to which (\ref{eq:ellipticSchroedinger})~is finite-band if~and only~if for
all $B\in\mathbb{C}$, every solution~$\Psi$ is meromorphic on~$E_{g_2,g_3}$
(multi-valuedness being allowed).

For example, when $q(x)\defeq\ell(\ell+1)x$, there is a {\em polynomial\/}
solution ${\mathcal{F}}_\ell(x;B;g_2,g_3)$ of~(\ref{eq:pseudohalphen}) that
satisfies conditions (i) and~(ii), of degree~$\ell$ in both $x$~and~$B$.
So the integer-$\ell$ Lam\'e equation is a finite-band Schr\"odinger
equation.  ${\mathcal{F}}_\ell(x;B;g_2,g_3)$~is called the $\ell$'th
Hermite--Halphen polynomial.  It may be written
$\hat{\mathcal{F}}_\ell(x;B;g_2,g_3)$ when it is normalized to be monic
in~$x$, rather than in~$B$ (see table~\ref{tab:hh}).

\begin{table}
\caption{Hermite--Halphen polynomials
\protect\mycite[table~A.2]{vanderWaall2002}}
\hfill
\begin{tabular}{ll}
\hline
$\ell$ & $\hat{\mathcal{F}}_\ell(x;B;g_2,g_3)$\\
\hline
$1$ & $x-B$\\
$2$ & $x^2 - \frac13Bx + (\frac19B^2 -\frac14g_2)$\\
$3$ & $x^3 -\frac15Bx^2 + \left(\frac2{75}B^2-\frac14g_2\right)x + (-
\frac1{225}B^3 +\frac1{15}Bg_2 - \frac14g_3)$\\
\hline
\end{tabular}
\hfill
\label{tab:hh}
\end{table}

In the case of a general rational potential~$q(x)$ which is finite-band,
let ${\cal F}(x;B)$ denote the specified solution
of~(\ref{eq:pseudohalphen}), and let its degree in~$B$ be denoted~$\ell$.
It~follows from manipulations parallel to those of Whittaker \& Watson that
the function~$\Psi$ on~$E_{g_2,g_3}$ defined by a formula containing an
indefinite integral,
\begin{equation}
\label{eq:hh}
\Psi(x,y;B,\nu)\defeq
\exp\int
\left[
\frac
{\frac12 \mathcal{F}^{\,\prime}(x;B)\,y - \nu}
{\mathcal{F}(x;B)}
\right]
\frac{\rd x}y,
\end{equation}
will be a solution of the Schr\"odinger
equation~(\ref{eq:ellipticSchroedinger}).  Here $B\in\mathbb{C}$ and
$\nu$~is a $B$-dependent but position-independent quantity, determined only
up~to sign, that can be computed by what Whittaker \& Watson term an
``interesting formula,''
\begin{equation}
{\nu}^2 = -\frac12\mathcal{F}\left(y\frac{\rd}{\rd
x}\right)^2\mathcal{F} + \left[\frac12 \left(y\frac{\rd}{\rd
x}\right)\mathcal{F}\right]^2 + [q(x)+B]\,\mathcal{F}_\ell^2.
\end{equation}
(It~is not obvious that the right-hand side is independent of the point
$(x,y)\in E_{g_2,g_3}$.)  It~is widely known \mycite{Smirnov2001} that
$\nu$~is identical to the coordinate~$\nu$ on the spectral
curve~$\Gamma_\ell$ defined by $\nu^2=\prod_{s=0}^{2\ell} (B-B_s)$, where
$\{B_s\}_{s=0}^{2\ell}$ are the band edges of the Schr\"odinger operator;
though no~really simple proof of this fact seems to have been published.
A~consequence of this is that the formula (\ref{eq:hh})~{\em
parametrizes\/} solutions of the elliptic-curve Schr\"odinger
equation~(\ref{eq:ellipticSchroedinger}) by
$(B,\nu)\in\Gamma_\ell\setminus\{(\infty,\infty)\}$.  As~defined, $\Psi$~is
called a Baker--Akhiezer function~\mycite{Krichever90orig}.

Consider now the special case of the integer-$\ell$ Lam\'e
equation~(\ref{eq:ellipticlame}).  In~this case the function~$\Psi$
computed by~(\ref{eq:hh}) from the Hermite--Halphen polynomials ${\cal
F}={\cal F}_\ell(x;B;g_2,g_3)$ is in~fact an Hermite--Halphen solution of
the Lam\'e equation, reexpressed in~terms of the elliptic curve
coordinates~$(x,y)$.  One can write $\Psi=\Psi_\ell^\pm(x,y;B;g_2,g_3)$,
where the superscript `$\pm$'~refers to the ambiguity in the sign of
$\nu=\nu(B)$.  If $\nu\neq0$, the two solutions~$\Psi_\ell^\pm$ are
distinct.  They are path-multiplicative, since they are exponentials of
anti-derivatives of rational functions on~$E_{g_2,g_3}$.

It should be noted that the Hermite--Halphen polynomials are not merely a
tool for generating the solutions~$\Psi_\ell^\pm$ of the Lam\'e equation.
They are algebraically interesting in their own right.
\myciteasnoun[figures~1,2]{Klein1892} supplies a sketch of the real portion
of the curve $\mathcal{F}_\ell(x;B)=0$ when $\ell=5,6$, showing how
when~$\ell\ge4$, $B$~is a band edge only if $\mathcal{F}_\ell(x;B)$,
regarded as a polynomial in~$x$, has a double root.

The relevance of the fundamental multi-valued function~$\Phi$ can now be
explained.  It~follows from~(\ref{eq:hh}), and the fact that ${\cal
F}_1=B-x$ (see table~\ref{tab:hh}), that
\begin{equation}
\label{eq:phisolns}
\Psi_1^\pm(x,y;B;g_2,g_3) = \Phi(x,y;B,\pm\sqrt{4B^3-g_2B-g_3}).
\end{equation}
That~is, if $(x_0,y_0)\in E_{g_2,g_3}$ is `above' $x_0=B$, then
$\Phi(\cdot,\cdot;x_0,y_0)$ will be a solution of the $\ell=1$ Lam\'e
equation in the form~(\ref{eq:ellipticlame}).  There are two such points,
related by $y_0$ being negated, unless $4B^3-g_2B-g_3=0$, i.e., unless
$B=e_1,e_2,e_3$, in which case $y_0=0$ is the only possibility.  These are
the three band-edge values of~$B$ for~$\ell=1$.

It is not difficult to show that the $\ell=1$ Hermite--Halphen
solutions~(\ref{eq:phisolns}) are identical to the
solutions~(\ref{eq:jacobiPhi}), though they are expressed as functions of
the variable $(x,y)\in E_{g_2,g_3}$ rather than the original independent
variable $\alpha\in\mathbb{C}$.  The parametrizing point~$(x_0,\pm y_0)\in
E_{g_2,g_3}$ corresponds to the parameter $\pm\alpha_0\in\mathbb{C}$
of~(\ref{eq:jacobiPhi}).  These solutions are clearly easier to formulate
in the elliptic curve context.

For any integer~$\ell$, the Lam\'e dispersion relation can be computed
numerically from~(\ref{eq:hh}) by calculating the multiplier arising from
the path of integration winding around~$E_{g_2,g_3}$.  However,
(\ref{eq:hh})~is not adapted to symbolic computation.  By expanding the
integrand in partial fractions one can derive the remarkable formula
\begin{equation}
\Psi_\ell^\pm(x,y;B;g_2,g_3) = \prod_{r=1}^\ell \Phi(x,y;x_r,y_r^\pm),
\end{equation}
where $\{(x_r,y_r^\pm)\}_{r=1}^\ell$ are points on~$E_{g_2,g_3}$ above
$\{x_r\}_{r=1}^\ell$, the $B$-dependent roots of the degree-$\ell$
polynomial~$\mathcal{F}_\ell(x;B;g_2,g_3)$.
(Cf.~\myciteasnoun[\S\,23.7]{Whittaker27}.)  Unfortunately, when $\ell\ge5$ the
roots $\{x_r\}_{r=1}^\ell$ cannot be computed in~terms of radicals.  This
reduction to degree-$1$ solutions is less computationally tractable than
the one that will be provided by the Hermite--Krichever Ansatz.

\section{Finite families of Lam\'e equation solutions}
\label{sec:families}

The solutions of the integer-$\ell$ Lam\'e equation include the Lam\'e
polynomials, which are the traditional band-edge solutions.  In the
Jacobi-form context they are periodic or anti-periodic functions
on~$[0,2K]$, with Floquet multiplier $\xi=\pm1$, respectively.  There are
exactly $2\ell+1$ values of the spectral parameter $B\in\mathbb{C}$, i.e.,
of the energy~$E$, for which a Lam\'e polynomial may be constructed, the
counting being up~to multiplicity.  By definition these are the roots of
the spectral polynomial $\mathsf{L}_\ell(B;g_2,g_3)$.

As functions on the curve~$E_{g_2,g_3}$, the Lam\'e polynomials are single
or double-valued and are essentially polynomials in the coordinates~$x,y$.
(In~the Weierstrassian context $\wp,\wp'$ substitute for~$x,y$.)  However,
no~fully satisfactory table of the Lam\'e polynomials or the Lam\'e
spectral polynomials has yet been published.
\myciteasnoun[\S\,23.42]{Whittaker27} refer to a list of
\myciteasnoun{Guerritore09} that covers~$\ell\le10$.  Sadly, although he
produced it as a {\em dissertazione di laurea\/} at the University of
Naples, most of his results on~$\ell\ge5$ are incorrect.  This has long
been known \mycite{Strutt67}, but his paper is still occasionally cited for
completeness \mycite{Gesztesy2003}.  \myciteasnoun[\S\,9.3.2]{Arscott64} gives
a brief table of the Jacobi-form Lam\'e polynomials, covering only
$\ell=1,2,3$.  His table is correct, with a single misprint
\mycite{Fernandez2000,Finkel2000}.  But its brevity has been misinterpreted.
An erroneous belief has arisen that when $\ell\ge4$, the Lam\'e polynomial
coefficients and band edge energies cannot be expressed in~terms of
radicals.  This sets~in only when~$\ell\ge8$.

Due to these confusions, in this section we tabulate the Lam\'e polynomials
and the spectral polynomials $\mathsf{L}_\ell(B;g_2,g_3)$.  Both are
computed from coefficient recurrence relations.  We~supply such relations
and tables of spectral polynomials for the twisted and theta-twisted Lam\'e
polynomials, as~well.  The number of values of $B\in\mathbb{C}$ for which
the latter two sorts of solution exist, i.e., the degrees of their spectral
polynomials, will be given.  All three sorts of solution will play a role
in Theorem~L\null.  In~fact, all will be special cases of the solutions
constructed for arbitrary~$B$ by the Hermite--Krichever Ansatz.

When $\ell\ge2$, many of the spectral polynomials will have degenerate
roots if $g_2,g_3\in\mathbb{C}$ are appropriately chosen.  This means that,
for~example, a~pair of the $2\ell+1$ band edge energies can be made to
coincide by moving the modular parameter~$m\in\mathbb{C}\setminus\{0,1\}$
to one of a finite set of complex values.  We~indicate how to calculate
these, or the corresponding values of Klein's absolute invariant
$J=g_2^3/(g_2^3-27g_3^2)\in\mathbb{C}$.

$J$~is the more fundamental parameter, in algebraic geometry at~least,
since two elliptic curves are isomorphic (birationally equivalent) if and
only if they have the same value of~$J$.  The $m\mapsto J$
correspondence~(\ref{eq:mtoJ}) maps $\mathbb{C}\setminus\{0,1\}$
onto~$\mathbb{C}$, and it also maps $m\in\mathbb{R}\setminus\{0,1\}$
(in~fact, $m\in(0,\frac12]$) onto $J\in[1,\infty)$.  Formally it is
$6$-to-$1$.  Each value of~$J$ corresponds to six values of~$m$, with the
exception of $J=0$ (i.e., $g_2=0$), which corresponds to $m=\frac12\pm
\frac{\sqrt3}2\ri$, and $J=1$ (i.e., $g_3=0$), which corresponds to
$m=-1,\frac12,2$.  Elliptic curves with $J=0,1$ are called {\em
equianharmonic\/} and {\em lemniscatic\/}, respectively \mycite[\S\S\,18.13
and~18.15]{Abramowitz65}.  Any equianharmonic curve has a triangular period
lattice, with $\omega'/\omega=\re^{\pm 2\pi \ri/3}$, and any lemniscatic
curve has a square period lattice, with $\omega'/\omega=\pm \ri$.

\subsection{Lam\'e polynomials}

The Lam\'e polynomials are classified into species $1,2,3,4$
\mycite[\S\,23.2]{Whittaker27}.  This is appropriate for some forms of the
Lam\'e equation, but for the elliptic-curve algebraic form, a more
structured classification scheme is better.

\begin{definition}
A solution of the Lam\'e equation~(\ref{eq:ellipticlame}) on the elliptic
curve~$E_{g_2,g_3}$ is said to be a Lam\'e polynomial of Type~I if it is
single-valued and of the form $C(x)$ or~$D(x)y$, where $C,D$ are
polynomials.  A~solution is said to be a Lam\'e polynomial of Type~II,
associated with the branch point~$e_\gamma$ of the curve ($\gamma=1,2,3$),
if it is double-valued and of the form $E(x)\sqrt{x-e_\gamma}$ or
$F(x)y/\sqrt{x-e_\gamma}$, where $E,F$ are polynomials.  The subtypes of
Types I and~II are species $1,4$ and~$2,3$, respectively.
\end{definition}

To determine necessary conditions on $\ell$ and~$B$ for there to be a
nonzero Lam\'e polynomial of each subtype, one may substitute the
corresponding expression ($C(x)$, etc.) into the Lam\'e
equation~(\ref{eq:ellipticlame}), and work~out a recurrence for the
polynomial coefficients.  This is similar to the approach of expanding in
integer or half-integer powers of~$x-e_\gamma$
\mycite[\S\,23.41]{Whittaker27}, though it leads to four-term rather than
three-term recurrences.  For the Type~I solutions at~least, the present
approach seems more natural, since they are not associated with any
singular point~$e_\gamma$.

\begin{table}
\caption{Lam\'e polynomials of Types~I,II}
\begin{tabular}{ll}
\hline
$\ell$ & Type~I solution: $C(x;B,g_2,g_3)$ or $D(x;B,g_2,g_3)\,y$\\
\hline
$1$ & --- \\
$2$ & $x-\frac16B$\\
$3$ & $y$\\
$4$ & $x^2-\frac1{14}Bx+(\frac1{280}B^2-\frac{3}{20} g_2)$\\
$5$ & $\left[x-\frac1{18}B\right]y$\\
$6$ & $x^3-\frac1{22}B x^2+(\frac1{792}B^2 -\frac{5}{24} g_2) x + (-\frac1{33264}B^3+\frac{13}{1584} B g_2-\frac17g_3)$\\
$7$ & $\left[x^2-\frac1{26}B x+\frac{1}{1144}B^2-\frac{5}{44} g_2\right]y$\\
$8$ & $x^4-\frac1{30}Bx^3+(\frac1{1560}B^2-\frac7{26}
g_2)x^2+(-\frac1{102960}B^3+\frac9{1144} B g_2-\frac2{11}
g_3)x$\\
    & $\quad{}+(\frac1{7413120}B^4-\frac7{51480} B^2 g_2+ \frac7{1320} B g_3+\frac7{624} g_2^2)$\\
\hline
\end{tabular}
\vskip0.15in				%add a blank line between sub-tables
\begin{tabular}{ll}
\hline
$\ell$ & Type~II solution: $E(x;B,e_\gamma,g_2,g_3)\sqrt{x-e_\gamma}$ or $F(x;B,e_\gamma,g_2,g_3)\,y/\sqrt{x-e_\gamma} $\\
\hline
$1$ & $\sqrt{x-e_\gamma}$\\
$2$ & $y/\sqrt{x-e_\gamma}$\\
$3$ & $\left[x + (-\frac1{10}B+\frac12e_\gamma)\right]\sqrt{x-e_\gamma}$\\
$4$ & $\left[x + (- \frac1{14}B-\frac12e_\gamma) \right]y/\sqrt{x-e_\gamma}$\\
$5$ & $\left[x^2 +(-\frac1{18}B+\frac12e_\gamma  ) x+(\frac1{504}B^2-\frac1{36}B e_\gamma+\frac{3}{8} e_\gamma^2-\frac{5}{28} g_2) \right]\sqrt{x-e_\gamma}$\\
$6$ & $\left[x^2 +(-\frac1{22}B-\frac12e_\gamma  ) x +(\frac1{792}B^2+\frac1{44}B e_\gamma -\frac18e_\gamma^2-\frac1{12}g_2)\right]y/\sqrt{x-e_\gamma}$\\
$7$ & $\left[x^3+(-\frac1{26}B + \frac12e_\gamma )x^2+(\frac1{1144}B^2 -\frac1{52}B e_\gamma +\frac38 e_\gamma^2 -\frac{21}{88} g_2 )x\right.$\\
    & $\quad\left.{}+(-\frac1{61776}B^3+\frac1{2288}B^2 e_\gamma-\frac3{208} B e_\gamma^2+\frac{493}{61776} B g_2-\frac{29}{704} e_\gamma g_2-\frac{145}{1728} g_3)\right]\sqrt{x-e_\gamma}$\\
$8$ & $\left[x^3+(-\frac1{30}B-\frac12e_\gamma )x^2+(\frac1{1560}B^2 +\frac1{60}B e_\gamma -\frac18e_\gamma^2 -\frac{15}{104} g_2 )x\right.$\\
    & $\quad\left.{}+(-\frac1{102960}B^3-\frac1{3120}B^2 e_\gamma+\frac1{240}B e_\gamma^2+\frac{127}{34320} B g_2+\frac{47}{832} e_\gamma g_2-\frac{51}{704} g_3)\right]y/\sqrt{x-e_\gamma}$\\
\hline
\end{tabular}
\label{tab:polys}
\end{table}

If $C(x)=\sum_j c_jx^j$, $D(x)=\sum_j d_jx^j$, $E(x)=\sum_j e_jx^j$, and
$F(x)=\sum_j f_jx^j$, substituting the expression for each species of
solution into~(\ref{eq:ellipticlame}) and equating the coefficients of
powers of~$x$ leads to the recurrence relations
\begin{align}
\label{eq:recurrence1}
\begin{split}
&(2j-\ell)(2j+\ell+1)\,c_j - B\,c_{j+1} \\
&\qquad {}- (j+2)(j+\tfrac32)g_2\,c_{j+2} - (j+2)(j+3)g_3\,c_{j+3} =0,
\end{split}
\\[\jot]
\label{eq:recurrence2}
\begin{split}
&(2j-\ell+3)(2j+\ell+4)\,d_j - B\,d_{j+1} \\
&\qquad {}- (j+2)(j+\tfrac52)g_2\,d_{j+2} - (j+2)(j+3)g_3\,d_{j+3} =0,
\end{split}
\displaybreak[0]\\[\jot]
\label{eq:recurrence3}
\begin{split}
&(2j-\ell+1)(2j+\ell+2)\,e_j + [(4j+5)e_\gamma - B]\,e_{j+1}\\
&\qquad {}+ [-(j+\tfrac52)g_2 + 4e_\gamma^2](j+2)\,e_{j+2} - (j+2)(j+3)g_3 \,e_{j+3} =0,
\end{split}
\\[\jot]
\label{eq:recurrence4}
\begin{split}
&(2j-\ell+2)(2j+\ell+3)\,f_j + [-(4j+7)e_\gamma - B]\,f_{j+1}  \\
&\qquad {}+ [-(j+\tfrac32)g_2 - 4e_\gamma^2](j+2)\,f_{j+2} - (j+2)(j+3)g_3 \,f_{j+3} =0.
\end{split}
\end{align}
It is easy to determine the integers~$\ell$ for which $C,D,E,F$ may be a
polynomial.

\begin{proposition}
If $\ell\ge1$ is odd, nonzero Type~I Lam\'e polynomials of the fourth
species and Type~II ones of the second species can in~principle be
constructed from these recurrence relations, with $\deg D=(\ell-3)/2$ and
$\deg E=(\ell-1)/2$ respectively.  {\rm(}The former assumes
$\ell\ge3$.\/{\rm)} If $\ell\ge2$ is even, nonzero Type~I Lam\'e polynomial
of the first species and Type~II ones of the third species can be
constructed similarly, with $\deg C=\ell/2$ and $\deg F=(\ell-2)/2$
respectively.
\end{proposition}

The coefficients in each Lam\'e polynomial are computed from the
appropriate recurrence relation by setting the coefficient of the highest
power of~$x$ to unity, and working downward.  Unless $B$~is specially
chosen, the coefficients of negative powers of~$x$ may be nonzero.  But by
examination, they will be zero if the coefficient of~$x^{-1}$ equals zero.

\begin{table}
\caption{Lam\'e spectral polynomials of Types~I,II}
\longcaption{Most of the ones with $\ell\ge5$ disagree with those published by
\protect\myciteasnoun{Guerritore09}.}
{\leavevmode\hfill
\begin{tabular}{l}
\begin{tabular}{ll}
\hline
$\ell$ & $L^{\rm I}_\ell(B;g_2,g_3)$\\
\hline
$1$ & $1$\\
$2$ & $B^2-3g_2$\\
$3$ & $B$\\
$4$ & $B^3-52g_2B + 560g_3$\\
$5$ & $B^2-27g_2$\\
$6$ & $B^4-294g_2B^2 + 7776g_3B + 3465g_2^2$\\
$7$ & $B^3-196g_2B + 2288g_3$\\
$8$ & $B^5-1044g_2B^3+48816g_3B^2+112320g_2^2B-4665600g_2g_3$\\
\hline
\end{tabular}\\
\\				%add a blank line between sub-tables
\begin{tabular}{ll}
\hline
$\ell$ & $L^{\rm II}_\ell(B;e_\gamma,g_2,g_3)$\\
\hline
$1$ & $B-e_\gamma$\\
$2$ & $B+3e_\gamma$\\
$3$ & $B^2-6e_\gamma B + (45e_\gamma^2 - 15g_2)$\\
$4$ & $B^2+10e_\gamma B + (-35e_\gamma^2 - 7g_2)$\\
$5$ & $B^3 -15 e_\gamma B^2  +(315 e_\gamma^2  -132  g_2)B +(\frac{675}4 e_\gamma g_2+ \frac{2835}4 g_3)$\\
$6$ & $B^3 +21 e_\gamma B^2 +(-189 e_\gamma^2  -84  g_2)B +(-\frac{3465}4 e_\gamma g_2+ \frac{4455}4 g_3)$\\
$7$ & $B^4 -28 e_\gamma B^3 +(1134 e_\gamma^2-574 g_2)B^2  +(3409  e_\gamma g_2  +8525  g_3)B$\\
    & $\quad{}+(-\frac{292383}4 e_\gamma^2 g_2-\frac{175175}4 e_\gamma g_3+22113 g_2^2)$\\
$8$ & $B^4 + 36 e_\gamma B^3 + (- 594 e_\gamma^2 - 414 g_2) B^2 + (- 9855
e_\gamma g_2 + 12285 g_3 ) B$\\
    & $\quad{} + (\frac{245025}4 e_\gamma^2
g_2 + \frac{552825}4 e_\gamma g_3 + 7425 g_2^2 )$\\
\hline
\end{tabular}
\end{tabular}
\hfill
}
\label{tab:spectral}
\end{table}

\begin{definition}
The Type-I Lam\'e spectral polynomial $L^{\rm I}_\ell(B;g_2,g_3)$ is the
polynomial monic in~$B$ which is proportional to the coefficient~$d_{-1}$
if $\ell$~is odd and $c_{-1}$ if $\ell$~is even.  (The former assumes
$\ell\ge3$; by convention $L^{\rm I}_1\defeq1$.)  The Type-II Lam\'e
spectral polynomial $L^{\rm II}_\ell(B;e_\gamma,g_2,g_3)$ is similarly
obtained from the coefficient~$e_{-1}$ if $\ell$~is odd and $f_{-1}$ if
$\ell$~is even.  Each spectral polynomial may be regarded as
$\prod_s[B-B_s(g_2,g_3)]$, resp.\ $\prod_s[B-B_s(e_\gamma,g_2,g_3)]$, where
the roots~$\{B_s\}$ are the values of~$B$ for which a Lam\'e equation
solution of the indicated type exists, counted with multiplicity.
\end{definition}

By examination, $N^{\rm I}_\ell\defeq\deg L^{\rm I}_\ell$ is $(\ell-1)/2$
if $\ell$~is odd and $\ell/2+1$ if $\ell$~is even; and $N^{\rm
II}_\ell\defeq\deg L^{\rm II}_\ell$ is $(\ell+1)/2$ if $\ell$~is odd and
$\ell/2$ if $\ell$~is even.  So as expected,
\begin{equation}
\label{eq:fullspectral}
\mathsf{L}_\ell(B;g_2,g_3) \defeq L^{\rm
I}_\ell(B;g_2,g_3)\prod_{\gamma=1}^3 L^{\rm II}_\ell(B;e_\gamma,g_2,g_3),
\end{equation}
the {\em full\/} Lam\'e spectral polynomial, has degree $N^{\rm I}_\ell + 3N^{\rm
II}_\ell = 2\ell+1$ in~$B$.

\begin{table}
\caption{Cohn polynomials of Types~I,II}
\begin{tabular}{l}
\begin{tabular}{ll}
\hline
$\ell$ & Type~I Cohn polynomial\\
\hline
$1$ & ---\\
$2$ & $J$\\
$3$ & ---\\
$4$ & $2^2\, 3^5\, J+5^2\, 7^2$\\
$5$ & $J$\\
$6$ & $2^4\, 3^2\, 5^2\, J^2+11\cdot 37\cdot 59 \, J-2^2\, 3^7$\\
$7$ & $2^4\, 3^5\, 5^2\, J+11^2\, 13^2$\\
$8$ & $J\,(2^{12}\, 3^5\, 5^2\, 7^2\, J^3+2^8\, 3^3\, 3664447\, J^2-2^4\,
3^2\cdot 397\cdot 364069\, J+{113}^5)$\\
\hline
\end{tabular}\\
\\				%add a blank line between sub-tables
\begin{tabular}{ll}
\hline
$\ell$ & Type~II Cohn polynomial\\
\hline
$1$ & ---\\
$2$ & ---\\
$3$ & $2^2\, J+1$\\
$4$ & $2^2\, 3^5\, J-5^3$\\
$5$ & $2^{14}\, 3^6\, J^3+2^9\, 3^3\, 5^2\, 109\, J^2-2^2\, 5^4\,17\cdot 151\, J+5^6\, 7^3$\\
$6$ & $2^{16}\, 3^6\, 5^2\, J^3+2^{11}\, 3^3\, 17\cdot 359\, J^2+2^4\, 57774169\, J+3^3\, 109^3$\\
$7$ & $2^{20}\, 3^{21}\, 5^6\, J^6+2^{16}\, 3^{17}\, 5^4\, 19\cdot 22307\,
J^5 + 2^{13}\, 3^{13}\, 5^5\, 22158751\, J^4$ \\
    & $\quad{}+2^9\, 3^6\, 5^2\, 1276543\cdot 414016613\, J^3- 2^4\, 3^5\, 5^2\,47202908378639011\, J^2$\\
    & $\quad{}+3^7\, {29}\cdot {41}\cdot {101}\cdot 895253\cdot 8050981\, J-2^5\, 5^6\, {11}^3\, {37}^3\, {113}^3$\\
$8$ & $2^{20}\, 3^9\, 5^2\, 7^2\, J^6+2^{16}\, 3^9\, 107\cdot 419\, J^5+2^{13}\, 3^7\, 12486499\, J^4-2^9\, 3^7\, 1171\cdot 10477\, J^3$\\
    & $\quad{}-2^4\, 3^4\, 11\cdot 47\cdot 91938173\, J^2+3^4\, 20593\cdot 844499\, J-2^5\, 7^3\, 13^3\, 29^3$\\
\hline
\end{tabular}
\end{tabular}
\label{tab:cohn}
\end{table}

It should be noted that $\prod_{\gamma=1}^3 L^{\rm
II}_\ell(B;e_\gamma,g_2,g_3)$, the full Type-II Lam\'e spectral polynomial,
is a function only of~$B;g_2,g_3$, since any symmetric polynomial
in~$e_1,e_2,e_3$ can be written in~terms of~$g_2,g_3$.  For~example,
$e_1e_2e_3=g_3/4$.  This is why $e_\gamma$~is absent on the left-hand side
of~(\ref{eq:fullspectral}).  When using the recurrences
(\ref{eq:recurrence1})--(\ref{eq:recurrence4}), one should also note that
$4e_\gamma^3-g_2e_\gamma-g_3=0$, so $e_\gamma^3=\frac14(g_2e_\gamma+g_3)$.
Any polynomial in~$e_\gamma,g_2,g_3$ can be reduced to one which is of
degree at~most~$2$ in~$e_\gamma$, much as any polynomial in~$x,y$ can be
reduced to one of degree at~most~$1$ in~$y$.

The Lam\'e polynomials of Types I and~II are listed in
table~\ref{tab:polys}, and the corresponding spectral polynomials in
table~\ref{tab:spectral}.  They replace the table of
\myciteasnoun{Guerritore09}, with its many unfortunate errors.  The spectral
polynomials with $\ell\le7$ were recently computed by a different technique
\mycite[table~A.3]{vanderWaall2002}.  The table of van~der Waall displays the
full Type~II spectral polynomials, rather than the more fundamental
$e_\gamma$-dependent polynomials $L^{\rm II}_\ell(B;e_\gamma,g_2,g_3)$.
The spectral polynomials can also be computed by the technique of
\myciteasnoun{Gesztesy95}, which employs the Weierstrassian counterpart of
the Ansatz used by Hermite in his solution of the Jacobi-form Lam\'e
equation~\mycite[\S\,23.71]{Whittaker27}.

The roots of the spectral polynomials are the energies~$B$ for which the
Lam\'e polynomials are solutions of the Lam\'e equation.  It~is clear that
when $\ell\ge9$, the Type~II energies cannot be expressed in~terms of
radicals, since the degree of the spectral polynomial will be $5$~or above.
When $\ell=8$ or~$\ell\ge10$, the Type~I energies cannot be so expressed.
These statements apply also to the coefficients of the Lam\'e polynomials,
which depend on~$B$.  So when $\ell\ge10$, the symbolic computation of the
Lam\'e polynomials is impossible, and when $\ell=8$ or~$9$, it~is possible
only in~part.  But when $g_2,g_3$ take~on special values, what would
otherwise be impossible may become possible.  For instance, when $g_3=0$
(the lemniscatic case, including $m=\frac12$), the quintic spectral
polynomial $L^{\rm I}_8(B;g_2,g_3)$ reduces to
$B^5-1044g_2B^3+112320g_2^2B$, the roots of which can obviously be
expressed in~terms of radicals.

In the context of the Jacobi form, the $2\ell+1$ values
$\{E_s(m)\}_{s=0}^{2\ell}$ for which a Lam\'e polynomial solution exists
can be thought of as the $2\ell+1$ branches of a spectral curve that lies
over over the triply-punctured sphere
$\mathbb{P}^1\setminus\{0,1,\infty\}$, the space of values of the modular
parameter~$m$.  \myciteasnoun{Turbiner89} showed that if $\ell\ge2$, this
curve has only four connected components, not~$2\ell+1$.  It~is now clear
why.  These are the Type~I component and the three Type~II components, one
associated with each point~$e_\gamma$.  Since each of the four is defined
by a polynomial in $E$ and~$m$, each can be extended to an algebraic curve
over~$\mathbb{P}^1$.  At~the values $m=0,1,\infty$, the four curves may
touch one another.  (See, e.g., \myciteasnoun[figure~3]{Li00}, for the
behavior of the real portions of the $\ell=1,2$ curves as~$m\to0,1$, and
\myciteasnoun{Alhassid83} for $\ell=3$.)  These three values of~$m$
correspond to two of $e_1,e_2,e_3$ coinciding, and the elliptic curve
$y^2=4\prod_{\gamma=1}^3(x-e_\gamma)$ becoming rational rather than
elliptic.  Level crossings of this sort are perhaps less interesting than
`intra-curve' ones.

In the present context, $E$~is replaced by the transformed energy~$B$, and
$m$ by the pair~$g_2,g_3$ or the Klein invariant~$J$, with $J=\infty$
corresponding to $m=0,1,\infty$.  It~is easy to determine which finite
values of~$J$ yield coincident values of~$B$.  One simply computes the
discriminants of the Type~I and full Type~II spectral polynomials, $L^{\rm
I}_\ell(B;g_2,g_3)$ and~$\prod_{\gamma=1}^3 L^{\rm
II}_\ell(B;e_\gamma,g_2,g_3)$.  Each discriminant is zero if and only if
there is a double root.  By using $J=g_2^3/(g_2^3-27g_3^2)$, one can
eliminate $g_2,g_3$ and obtain a polynomial equation for~$J$.  For each of
Types I and~II, there are coincident values of~$B$ if and only if $J$~is a
root of what may be called a {\em Cohn polynomial\/}.

In table~\ref{tab:cohn} the Cohn polynomials are listed.  Since the
coefficients are rather large integers that may have number-theoretic
significance, each is given in a fully factored form.  An interesting
feature of these polynomials is that none has a zero on the real
half-line~$[1,\infty)$.  Since $J\in[1,\infty)$ corresponds to $m\in(0,1)$,
the existence of such a zero would imply that for some $m\in(0,1)$, two of
the $2\ell+1$ band edges become degenerate.  That this cannot occur follows
from a Sturmian argument~\mycite[\S\,23.41]{Whittaker27}.  It~also follows
from the analysis of \myciteasnoun[\S\,3]{Gesztesy95}.

\begin{proposition}
\label{prop:cohnconj}
For any integer $\ell\ge1$, the degeneracies of the algebraic spectrum of
the Lam\'e operator, which comprises the $2\ell+1$ roots {\rm(}up~to
multiplicity\/{\rm)} of the spectral polynomial\/
$\mathsf{L}_\ell(B;g_2,g_3)$, are fully captured by the Cohn polynomials of
Types I and~II\null.  As~the parameters $g_2,g_3$ are varied, a pair of
roots will coincide, reducing the number of distinct roots from $2\ell+1$
to~$2\ell$, if and only if the Klein invariant~$J$ is a root of one of the
two Cohn polynomials; and there are no multiple coincidences.
\end{proposition}

\noindent
This proposition will be proved in~\S\,\ref{sec:HK}.  The following
conjecture is based on a close examination of the spectral and Cohn
polynomials, for all $\ell\le25$.

\begin{conjecture}
\mbox{}
\begin{enumerate}
\item As a polynomial in~$J$ with integer coefficients, no~Cohn polynomial
has a nontrivial factor, except for the Type~I Cohn polynomials with
$\ell\equiv2\pmod3$, each of which is divisible by~$J$.  {\rm(}These
factors of~$J$ are visible in table\/~{\rm\ref{tab:cohn}.}{\rm)}
\item If $N_\ell^{\rm I}$ and~$N_\ell^{\rm II}$ denote the degrees of the
spectral polynomials of Types I,II, which are given above, then the Cohn
polynomials of Types I,II have degrees $\left\lfloor ({N_\ell^{\rm I}}^2 -
N_\ell^{\rm I} + 4)/6 \right\rfloor$ and $N_\ell^{\rm II}(N_\ell^{\rm
II}-1)/2$, respectively.
\end{enumerate}
\label{conj:cohnconj}
\end{conjecture}

\noindent
The conjectured degree formulas constitute a conjecture as to the number of
points in elliptic moduli space (elliptic curve parameter space), labelled
by~$J$, at which the $2\ell+1$ distinct energies in the algebraic spectrum
are reduced to~$2\ell$.  For~example, $N_3^{\rm I},N_3^{\rm II}$ equal
$1,2$, so when $\ell=3$ the Cohn polynomials of Types I,II have degrees
$0,1$ respectively.  That~is, if $\ell=3$ there is no Type-I polynomial,
and the Type-II one is linear in the invariant~$J$.  According to
table~\ref{tab:cohn} it equals $4J+1$.  A~nondegeneracy condition
equivalent to the linear condition $4J+1\neq0$ was previously worked~out by
\myciteasnoun[\S\,6.6]{Treibich94}, namely $\prod_{\gamma=1}^3
(5g_2-12e_\gamma^2)\neq0$.

The remarks regarding extra $J$~factors amount to a conjecture that in the
equianharmonic case $m=\frac12\pm \frac{\sqrt3}2\ri$ (i.e., $J=0$
or~$g_2=0$), there are only $2\ell$ distinct energies if and only if
$\ell\equiv2\pmod3$.  For those values of~$\ell$, the double energy
eigenvalue is evidently located at~$B=0$.  It should be mentioned that when
$J=0$ and $\ell\equiv0\pmod3$, there is also an eigenvalue at~$B=0$, but it
is a simple one.

A~periodicity of length~$3$ in~$\ell$ is present in the equianharmonic case
of a third-order equation resembling~(\ref{eq:pseudohalphen}), now called
the Halphen equation \mycite[pp~571--4]{Halphen1886}.  By the preceding, a
similar periodicity appears to be present in the equianharmonic case of the
Lam\'e equation.  This was not previously realized.

\subsection{Twisted Lam\'e polynomials}

The twisted Lam\'e polynomials are exponentially modified Lam\'e
polynomials.  They will play a major role in Theorem~L and in the
hyperelliptic reductions following from the Hermite--Krichever Ansatz, but
they are of independent interest.

\begin{definition}
A solution of the Lam\'e equation~(\ref{eq:ellipticlame}) on the elliptic
curve~$E_{g_2,g_3}$ is said to be a twisted Lam\'e polynomial, of Type~I or
Type~II associated with the point~$e_\gamma$, if it has one of the two
forms
\begin{displaymath}
\genfrac\{\}{0pt}{}{C(x)+D(x)\,y}{E(x)\sqrt{x-e_\gamma}+F(x)\,y/\sqrt{x-e_\gamma}}\times\exp\left[\kappa\int\frac{\rd x}{y}   \right],
\end{displaymath}
with $\kappa\in\mathbb{C}$ nonzero.  Here $C,D,E,F$ are polynomials.
\end{definition}

On the level of Fuchsian differential equations, there is little to
distinguish between twisted Lam\'e polynomials and ordinary Lam\'e
polynomials, which are simply twisted polynomials with $\kappa=0$.
A~function $\Psi(x,y)=\widehat\Psi(x,y)\exp\left[\kappa\int \rd x/y\right]$
will be a solution of the Lam\'e equation~(\ref{eq:ellipticlame}) if and
only if $\widehat\Psi$ satisfies
\begin{equation}
\label{eq:newlame}
\left\{ \left(y\frac{\rd}{\rd x}\right)^2 + 2\kappa\left(y\frac{\rd}{\rd
x}\right) - \left[\ell(\ell+1)x + B-\kappa^2\right] \right\}\widehat\Psi=0.
\end{equation}
This is a Fuchsian equation on~$E_{g_2,g_3}$ that generalizes but strongly
resembles~(\ref{eq:ellipticlame}).  It~has a single regular singular point,
$(x,y)=(\infty,\infty)$, and its characteristic exponents there are
$\{-\ell,\ell+1\}$.  Like~$B$, $\kappa$~is an accessory parameter that does
not affect the exponents.  The values $(B,\kappa)$ for which a twisted or
conventional Lam\'e polynomial solution of~(\ref{eq:ellipticlame}) exists
can be viewed as the points in a two-dimensional accessory parameter space
at which (\ref{eq:newlame})~has single or double-valued solutions.

If $C(x)=\sum_j c_jx^j$, $D(x)=\sum_j d_jx^j$, $E(x)=\sum_j e_jx^j$, and
$F(x)=\sum_j f_jx^j$, substituting the expression for each type of twisted
polynomial solution into~(\ref{eq:ellipticlame}) and equating the
coefficients of powers of~$x$ yields the coupled pairs of recurrences
\begin{align}
\label{eq:twistedrecurrence1}
\begin{split}
&(2j-\ell)(2j+\ell+1)\,c_j +(\kappa^2 - B)\,c_{j+1} \\
&\qquad {}-(j+2)(j+\tfrac32)g_2\,c_{j+2} - (j+2)(j+3)g_3\,c_{j+3}\\
&\qquad {}+2\kappa\,(4j+2)\,d_{j-1} - 2\kappa\,(j+\tfrac32)g_2\,d_{j+1}\\
&\qquad {}-2\kappa\,(j+2)g_3\,d_{j+2} =0,
\end{split}
\\[\jot]
\label{eq:twistedrecurrence2}
\begin{split}
&(2j-\ell+3)(2j+\ell+4)\,d_j +(\kappa^2 - B)\,d_{j+1} \\
&\qquad {}-(j+2)(j+\tfrac52)g_2\,d_{j+2} - (j+2)(j+3)g_3\,d_{j+3}\\
&\qquad {}+2\kappa\,(j+2)\,c_{j+2}=0;
\end{split}
\displaybreak[0]\\[10pt]
\label{eq:twistedrecurrence3}
\begin{split}
&(2j-\ell+1)(2j+\ell+2)\,e_j + [(4j+5)e_\gamma +\kappa^2- B]\,e_{j+1}\\
&\qquad {}+[-(j+\tfrac52)g_2 + 4e_\gamma^2](j+2)\,e_{j+2} - (j+2)(j+3)g_3 \,e_{j+3}\\
&\qquad {}+2\kappa\,(4j+4)\,f_j + 2\kappa\,(4j+6)e_\gamma\,f_{j+1}\\
&\qquad {}+2\kappa\,(4j+8)(e_\gamma^2-\tfrac14g_2)\,f_{j+2}=0,
\end{split}
\\[\jot]
\begin{split}
\label{eq:twistedrecurrence4}
&(2j-\ell+2)(2j+\ell+3)\,f_j + [-(4j+7)e_\gamma + \kappa^2 - B]\,f_{j+1}\\
&\qquad {}+[-(j+\tfrac32)g_2 - 4e_\gamma^2](j+2)\,f_{j+2} - (j+2)(j+3)g_3
\,f_{j+3}\\
&\qquad {}+2\kappa\,(j+\tfrac32)\,e_{j+1} - 2\kappa\,(j+2)e_\gamma\,e_{j+2}=0.
\end{split}
\end{align}
It is easy to determine the maximum value of the exponent~$j$ in each
of~$C,D,E,F$.
\begin{proposition}
Nonzero twisted Lam\'e polynomials of Type~I\/ {\rm(}if~$\ell\ge3${\rm)}
and of Type~II\/ {\rm(}if~$\ell\ge2${\rm)} can in principle be constructed
from these recurrence relations.  If $\ell$~is odd, resp.\ even, then $\deg
C,D;E,F$ are $(\ell-1)/2,(\ell-3)/2;(\ell-1)/2,(\ell-3)/2$, resp.\
$\ell/2,(\ell-4)/2;\ell/2-1,\ell/2-1$.  The coefficients are computed by
setting the coefficient of the highest power of~$x$ to unity in $D$
and~$E$, resp.\ $C$ and~$F$, and working downward.
\end{proposition}

Unless $B,\kappa$ are specially chosen, the coefficients of negative powers
of~$x$ may be nonzero.  But by examination, they will be zero if the
coefficients of~$x^{-1}$ in $C$ and~$D$ (for Type~I), or $E$ and~$F$ (for
Type~II), equal zero.  $c_{-1}=0$, $d_{-1}=0$, and $e_{-1}=0$, $f_{-1}=0$,
are coupled polynomial equations in~$B,\kappa$, and their solutions may be
computed by polynomial elimination, e.g., by computing resultants.  A~minor
problem is the proper handling of the case~$\kappa=0$, in which
(\ref{eq:twistedrecurrence1})--(\ref{eq:twistedrecurrence4}) reduce to
(\ref{eq:recurrence1})--(\ref{eq:recurrence4}).  If $\ell$~is odd, resp.\ 
even, then $c_{-1}$ and~$f_{-1}$, resp.\ $d_{-1}$ and~$e_{-1}$, turn~out to
be divisible by~$\kappa$.  By~dividing the appropriate equations
by~$\kappa$ before solving each pair of coupled equations, the spurious
$\kappa=0$ solutions can be eliminated.

\begin{definition}
The Type-I twisted Lam\'e spectral polynomial $Lt^{\rm I}_\ell(B;g_2,g_3)$
is the polynomial monic in~$B$ which is proportional to the resultant of
$c_{-1},d_{-1}$ with respect to~$\kappa$, with $\kappa$~factors removed as
indicated.  (This assumes $\ell\ge3$; by convention $Lt^{\rm I}_1=Lt^{\rm
I}_2\defeq1$.)  The Type-II twisted Lam\'e spectral polynomial $Lt^{\rm
II}_\ell(B;e_\gamma,g_2,g_3)$ is similarly obtained from $e_{-1},f_{-1}$.
(This assumes $\ell\ge2$; by convention ${Lt^{\rm II}_1\defeq1}$.)  Each
twisted spectral polynomial may be regarded as $\prod_s[B-B_s(g_2,g_3)]$,
resp.\ $\prod_s[B-B_s(e_\gamma,g_2,g_3)]$, where the roots~$\{B_s\}$ are
the values of~$B$ for which a Lam\'e equation solution of the specified
type exists, counted with multiplicity.
\end{definition}

\begin{table}
\caption{Twisted Lam\'e spectral polynomials of Types~I,II}
\begin{tabular}{ll}
\hline
$\ell$ & $Lt^{\rm I}_\ell(B;g_2,g_3)$\\ 
\hline 
$1$ & $1$\\ $2$ & $1$\\ $3$
& $B^2-\frac{75}4g_2$\\ $4$ & $B^3-\frac{343}4g_2B-\frac{1715}2g_3$\\ $5$ &
$B^6-\frac{897}2g_2B^4-\frac{19845}2g_3B^3+\frac{546993}{16}g_2^2B^2+\frac{5893965}8g_2g_3B$\\
& $\quad{}+(\frac{4100625}4g_2^3-\frac{506345175}{16}g_3^2)$\\ $6$ &
$B^8-\frac{2751}2g_2 B^6-\frac{181521}2
g_3B^5+\frac{3407481}{16}g_2^2B^4+\frac{164862621}8g_2g_3B^3$\\ &
$\quad{}+(\frac{677951505}{16}g_2^3-\frac{15273476559}{16}g_3^2)B^2-\frac{3362086035}8g_2^2g_3B$\\
& $\quad{}+ (-\frac{15980285475}{4}g_2^4 + \frac{1664232587325}{16} g_2
g_3^2)$\\ $7$ & $B^{12} - 4186 g_2 B^{10} - \frac{1048223}2 g_3 B^9 +
\frac{17433633}8 g_2^2 B^8 + \frac{3510785355}8 g_2 g_3 B^7$\\ &
$\quad{}+(\frac{4590448625}4 g_2^3 - \frac{59437238487}2 g_3^2) B^6 -
\frac{3146848477773}{32} g_2^2 g_3 B^5$\\ & $\quad{}+(-
\frac{239496271862939}{256} g_2^4 + \frac{349377693363699}{16} g_2 g_3^2
)B^4$\\ & $\quad{}+ (\frac{2167403005460693}{128} g_2^3 g_3 -
\frac{10531741687878125}{32} g_3^3) B^3$\\ & $\quad{}+
(\frac{1196552376313749}8 g_2^5 - \frac{243386984019562383}{64} g_2^2
g_3^2) B^2$\\ & $\quad{}+ (-\frac{570084251356448829}{128} g_2^4
g_3+\frac{15458554942852896875}{128} g_2 g_3^3) B$\\ &
$\quad{}+(-\frac{1649721227262688125}{256} g_2^6
+\frac{16766233150463677881}{128} g_2^3 g_3^2
+\frac{285799721595172159375}{256} g_3^4)$\\ $8$ & $B^{15} -10188g_2B^{13}
-\frac{4944861}2g_3B^{12} +\frac{48623733}8g_2^2B^{11}
+\frac{33098210361}8g_2g_3B^{10}$\\ &
$\quad{}+(\frac{210211163145}8g_2^3-\frac{1634908193451}4g_3^2)B^9
-\frac{46667883177495}{32}g_2^2g_3B^8$\\ &
$\quad{}+(-\frac{9879747455405475}{256}g_2^4
+\frac{13114846350610875}{16}g_2g_3^2 )B^7$\\ &
$\quad{}+(\frac{27951449759004375}{128}g_2^3g_3-\frac{268903562388069375}{32}g_3^3)B^6$\\
&
$\quad{}+(\frac{1268095592996251875}{64}g_2^5-\frac{16057777970613965625}{32}g_2^2g_3^2)B^5$\\
& $\quad{}+(-\frac{67310182108529184375}{128}g_2^4g_3
+\frac{1998126475855699190625}{128}g_2g_3^3)B^4$\\ &
$\quad{}+(-\frac{1247302375822866515625}{256}g_2^6
+\frac{15203913100824300328125}{128}g_2^3g_3^2
+\frac{83437068242769811171875}{256}g_3^4)B^3$\\ &
$\quad{}+(\frac{2476060022819411015625}{16}g_2^5g_3
-\frac{67800902314274734921875}{16}g_2^2g_3^3)B^2$\\ &
$\quad{}+(\frac{8803296317899887890625}{16}g_2^7
-\frac{100408533824565875390625}8g_2^4g_3^2
-\frac{1003123416299777251171875}{16}g_2g_3^4)B$\\ &
$\quad{}+(\frac{14047813273244501953125}2g_2^6g_3
-305679140836374550781250g_2^3g_3^3$\\ &
$\qquad\quad{}+\frac{6263785284763018974609375}2g_3^5)$\\
\hline
\end{tabular}
\vskip0.15in
\begin{tabular}{ll}
\hline
$\ell$ & ${Lt}^{\rm II}_\ell(B;e_\gamma,g_2,g_3)$\\
\hline
$1$ & $1$\\
$2$ & $B-6e_\gamma$\\
$3$ & $B^2 -15e_\gamma B+(-225e_\gamma^2+\frac{75}4g_2)$\\
$4$ & $B^4-55 e_\gamma B^3+(-945 e_\gamma^2+\frac{539}4 g_2) B^2+(1960
e_\gamma g_2 +2450 g_3) B$\\
    & $\quad{}+(61740 e_\gamma^2 g_2-68600 e_\gamma g_3- 9261g_2^2)$\\
$5$ & $B^6-105 e_\gamma B^5+(-7245 e_\gamma^2+\frac{1707}2
      g_2)B^4+(\frac{16065}4 e_\gamma g_2+\frac{19845}4 g_3)B^3$\\
    & $\quad{}+(\frac{5077485}4
      e_\gamma^2 g_2-\frac{2679075}4 e_\gamma
      g_3-\frac{2419551}{16} g_2^2)B^2$\\
    & $\quad{}+(-\frac{56260575}4 e_\gamma^2g_3 + 1530900 e_\gamma g_2^2
      +\frac{54117315}{16} g_2 g_3)B$\\
    & $\quad{}+(-\frac{86113125}2 e_\gamma^2
      g_2^2+\frac{120558375}2 e_\gamma g_2 g_3+\frac{36905625}8
      g_2^3+\frac{506345175}8 g_3^2)$\\
$6$ & $B^9 -231 e_\gamma B^8 +(-21735 e_\gamma^2 +\frac{6699}2 g_2) B^7
+(\frac{255087}4 e_\gamma g_2 + \frac{610983}4 g_3) B^6$\\
    & $\quad{}+(\frac{72454095}4 e_\gamma^2 g_2 -\frac{58128273}4 e_\gamma g_3
      -\frac{38437119}{16} g_2^2) B^5$\\
    & $\quad{}+(\frac{118721295}4 e_\gamma^2 g_3  -\frac{7116417}2 e_\gamma g_2^2
-\frac{249847983}{16} g_2 g_3) B^4$\\
    & $\quad{}+(-\frac{8838982845}2 e_\gamma^2 g_2^2
+\frac{10241858781}2 e_\gamma g_2 g_3 +\frac{4499132715}8 g_2^3
+2338773426 g_3^2) B^3$\\
    & $\quad{}+(\frac{6418527885}2 e_\gamma^2 g_2 g_3 +\frac{8690105655}4 e_\gamma g_2^3 
-86375046681 e_\gamma g_3^2
-\frac{13550225535}8 g_2^2 g_3) B^2$\\
    & $\quad{}+(\frac{1189804891275}4 e_\gamma^2 g_2^3 +1089315875340 e_\gamma^2 g_3^2-\frac{1906302781845}4 e_\gamma g_2^2 g_3$\\
    & $\qquad\quad{}-\frac{783033988275}{16} g_2^4 -151293871575 g_2 g_3^2) B$\\
    & $\quad{}+(\frac{11649628111275}4 e_\gamma^2 g_2^2 g_3-335585994975 e_\gamma g_2^4 +4992697761975 e_\gamma g_2 g_3^2$\\
    & $\qquad\quad{}-\frac{15101369773875}{16} g_2^3 g_3 +12838365673650 g_3^3)$\\
\hline
\end{tabular}
\label{tab:twistedspectral}
\end{table}

The twisted Lam\'e spectral polynomials for $\ell\le8$ are listed in
table~\ref{tab:twistedspectral}.  The polynomials $Lt^{\rm II}_7$
and~$Lt^{\rm II}_8$ are omitted on~account of lack of space (their
respective degrees are $12$ and~$16$).  The following proposition will be
proved in~\S\,\ref{sec:HK}.

\begin{proposition}
\mbox{}
\begin{enumerate}
\item For any integer $\ell\ge3$, resp.\ $\ell\ge2$, there is a nontrivial
twisted spectral polynomial of Type~I, resp.\ Type~II.
\item For any integer $\ell\ge1$, $Nt^{\rm I}_\ell\defeq\deg Lt_\ell^{\rm
I}$ is $(\ell^2-1)/4$ if $\ell$~is odd and $\ell^2/4-1$ if $\ell$~is even;
and $Nt^{\rm II}_\ell\defeq\deg Lt_\ell^{\rm II}$ is $(\ell^2-1)/4$ if
$\ell$~is odd and $\ell^2/4$ if $\ell$~is even.
\end{enumerate}
\label{prop:twisted}
\end{proposition}

\noindent
So the {\em full\/} twisted Lam\'e spectral polynomial
$\mathsf{Lt}_\ell(B;g_2,g_3)$, which by~definition equals $Lt^{\rm
I}_\ell(B;g_2,g_3)\prod_{\gamma=1}^3 Lt^{\rm II}_\ell(B;e_\gamma,g_2,g_3)$,
will be of degree $Nt_\ell^{\rm I} +3Nt_\ell^{\rm II} = \ell^2-1$ in the
spectral parameter~$B$.  Like the ordinary degree-$(2\ell+1)$ spectral
polynomial~$\mathsf{L}_\ell$, $\mathsf{Lt}_\ell$~is a function of
$B;g_2,g_3$ only, because any symmetric polynomial in~$e_1,e_2,e_3$ can be
written in~terms of the invariants~$g_2,g_3$.

\subsection{Theta-twisted Lam\'e polynomials}

Lam\'e equation solutions of a third sort can be constructed for certain
values of the spectral parameter~$B$.  These are linear combinations, over
polynomials in the coordinate~$x$, of (i)~the multi-valued meromorphic
function $\Phi(x,y;x_0,y_0)$ parametrized by the point~$(x_0,y_0)\in
E_{g_2,g_3}\setminus\{(\infty,\infty)\}$, and (ii)~its derivative
\begin{equation}
\label{eq:phi1}
\Phi^{(1)}(x,y;x_0,y_0)\defeq \left(y\frac{\rd}{\rd x}\right)
\Phi(x,y;x_0,y_0) = \frac12\left( \frac{y+y_0}{x-x_0} \right)
\Phi(x,y;x_0,y_0).
\end{equation}
One way of seeing that $\Phi,\Phi^{(1)}$ are a natural basis is to note
that when $(x_0,y_0)=(e_\gamma,0)$, they reduce to
$\sqrt{x-e_\gamma},\,\frac12 y/\sqrt{x-e_\gamma}$.  So the class of
functions constructed from them will include the Lam\'e polynomials of
Type~II\null.

\begin{definition}
A solution of the Lam\'e equation~(\ref{eq:ellipticlame}) on the elliptic
curve~$E_{g_2,g_3}$ is said to be a theta-twisted Lam\'e polynomial if it
is of the form $\mathcal{A}(x)\Phi(x,y;x_0,y_0) +
2\mathcal{B}(x)\Phi^{(1)}(x,y;x_0,y_0)$, with $(x_0,y_0)\neq(e_\gamma,0)$
for $\gamma=1,2,3$.  Here $\mathcal{A},\mathcal{B}$ are polynomials, and
the innocuous `$2$' factor compensates for the `$\frac12$' factor
of~(\ref{eq:phi1}).
\end{definition}

If $\mathcal{A}(x)=\sum_j a_jx^j$ and $\mathcal{B}(x)=\sum_j b_jx^j$,
substituting this expression into~(\ref{eq:ellipticlame}) and equating the
coefficients of powers of~$x$ yields the coupled pair of recurrences
\begin{align}
\label{eq:thetarecurrence1}
\begin{split}
&(2j-\ell+1)(2j+\ell+2)\,a_j + [(4j+5)x_0 - B]\,a_{j+1} 
\\
&\qquad {}+ [-(j+\tfrac52)g_2 + 4x_0^2](j+2)\,a_{j+2} - (j+2)(j+3)g_3
\,a_{j+3}\\
&\qquad {} -2y_0(4j+6)\,b_{j+1} - 4x_0y_0(j+2)\,b_{j+2}=0,
\end{split}
\\[\jot]
\label{eq:thetarecurrence2}
\begin{split}
&(2j-\ell+2)(2j+\ell+3)\,b_j + [-(4j+7)x_0 - B]\,b_{j+1}  
\\
&\qquad {}+ [-(j+\tfrac32)g_2 - 4x_0^2](j+2)\,b_{j+2} - (j+2)(j+3)g_3
\,b_{j+3}\\
&\qquad {} + y_0(j+2)\,a_{j+2}=0.
\end{split}
\end{align}

\begin{proposition}
If $\ell\ge4$, nonzero theta-twisted polynomials can in~principle be
computed from these recurrences.  If $\ell$~is odd, resp.\ even, then
$\deg\mathcal{A},\mathcal{B}$ are ${(\ell-1)}/2,\allowbreak(\ell-5)/2$,
resp.\ $\ell/2-2,\ell/2-1$.  The coefficients are computed by setting the
coefficient of the highest power of~$x$ in $\mathcal{A}$,
resp.~$\mathcal{B}$, to unity, and working downward.
\end{proposition}

Unless $B$ and the point $(x_0,y_0)$ are specially chosen, the coefficients
of negative powers of~$x$ may be nonzero.  But by examination, they will be
zero if the coefficients of~$x^{-1}$ in $\mathcal{A}$ and~$\mathcal{B}$ are
both zero.  $a_{-1}=0$, $b_{-1}=0$ are equations in~$B;x_0,y_0$.  Together
with the identity $y_0^2=4x_0^3-g_2x_0-g_3$, they make~up a set of three
polynomial equations for these three unknowns.  This system may be solved
by polynomial elimination.  For~example, to obtain a single polynomial
equation for~$B$ (involving $g_2,g_3$ of~course), one may eliminate $y_0$
from $a_{-1}=0$, $b_{-1}=0$ by computing their resultants against the third
equation; and then eliminate~$x_0$.  Alternatively, a Gr\"obner basis
calculation may be performed \mycite{Brezhnev2000}.

Irrespective of which procedure is followed, there is a minor problem: the
handling of the improper case $(x_0,y_0)=(e_\gamma,0)$, in which
(\ref{eq:thetarecurrence1})--(\ref{eq:thetarecurrence2}) reduce to
(\ref{eq:recurrence3})--(\ref{eq:recurrence4}).  If $\ell$~is odd, resp.\
even, then the left-hand side of the equation $b_{-1}=0$, resp.\
$a_{-1}=0$, turns~out to be divisible by~$y_0$.  By~dividing the
appropriate equation by~$y_0$ before eliminating $x_0,y_0$, the spurious
solutions with $y_0=0$ can be eliminated.

\begin{definition}
The theta-twisted Lam\'e spectral polynomial $L\theta_\ell(B;g_2,g_3)$ is
the polynomial monic in~$B$ which is obtained by eliminating $x_0,y_0$ from
the equations $a_{-1}=0$, $b_{-1}=0$, with $y_0$~factors removed as
indicated.  (This assumes $\ell\ge4$; by convention
$L\theta_1=L\theta_2=L\theta_3\defeq1$.)  Each theta-twisted spectral
polynomial may be regarded as $\prod_s[B-B_s(g_2,g_3)]$, where the
roots~$\{B_s\}$ are the values of~$B$ for which a theta-twisted Lam\'e
polynomial exists, counted with multiplicity.
\end{definition}

\begin{table}
\caption{Theta-twisted Lam\'e spectral polynomials}
\hfill
\begin{tabular}{ll}
\hline
$\ell$ & ${L\theta}_\ell(B;g_2,g_3)$\\
\hline
$1$ & $1$\\
$2$ & $1$\\
$3$ & $1$\\
$4$ & $B^2-\frac{196}3g_2$\\
$5$ & $B^3-\frac{1053}4g_2B-\frac{25515}4g_3$\\
$6$ & $B^6-\frac{4599}4 g_2 B^4 -\frac{120285}2 g_3 B^3 +160083 g_2^2
B^2+\frac{20376279}2 g_2 g_3 B$\\
    & $\quad{}+(-\frac{576357606}4g_3^2 +96850215g_2^3)$\\
$7$ & $B^8 -\frac{19565}6 g_2 B^6 -\frac{832843}2 g_3 B^5
+\frac{26047931}{48} g_2^2 B^4 +\frac{4205970769}{24} g_2 g_3 B^3$\\
    & $\quad{}+(\frac{37048456991}{48} g_2^3 -\frac{204966441251}{16} g_3^2) B^2
+\frac{8684628953}6 g_2^2 g_3 B$\\
    & $\quad{}+(-\frac{552623218875}4 g_2^4
+\frac{43902444356771}{12} g_2g_3^2)$\\
$8$ & $B^{12} -\frac{18063}2g_2B^{10} -\frac{4067739}2g_3B^9
+\frac{73174185}{16}g_2^2B^8 +\frac{22697632971}8g_2g_3B^7$\\
    &$\quad{}+\frac{28431}{16}(10247115 g_2^3 - 167402573 g_3^2)B^6
-\frac{1385229823965}2g_2^2g_3B^5$\\
    &$\quad{}+\frac{492075}4 (- 164228833 g_2^4 + 3606494307 g_2g_3^2)B^4$\\
    &$\quad{}+63969750000\, (2175 g_2^3 g_3 - 101062 g_3^3)B^3$\\
    &$\quad{}+98415000\, (62738863 g_2^5 - 1656031845 g_2^2g_3^2)B^2$\\
    &$\quad{}+921164400000\,(-256036 g_2^4g_3 + 7098507 g_2g_3^3)B$\\
    &$\quad{}+ 19683000000\,(-35153041 g_2^6 +669725199 g_2^3  g_3^2 +7578832716 g_3^4)$\\
\hline
\end{tabular}
\hfill
\label{tab:thetatwistedspectral}
\end{table}

The theta-twisted Lam\'e spectral polynomials for $\ell\le8$ are listed in
table~\ref{tab:thetatwistedspectral}.  The following proposition will be
proved in~\S\,\ref{sec:HK}.

\begin{proposition}
\mbox{}
\begin{enumerate}
\item For any integer $\ell\ge4$ there is a nontrivial theta-twisted spectral
polynomial~$L\theta_\ell$.
\item For any integer $\ell\ge2$, $N\theta_\ell\defeq\deg L\theta_\ell$ is
$(\ell+1)(\ell-3)/4$ if $\ell$~is odd and $\ell(\ell-2)/4$ if $\ell$~is
even.
\end{enumerate}
\label{prop:thetatwisted}
\end{proposition}

\section{The Hermite--Krichever Ansatz}
\label{sec:HK}

The Hermite--Krichever Ansatz is a tool for solving any Schr\"odinger-like
differential equation, not necessarily of second order, with coefficient
functions that are elliptic.  Such an equation should ideally have one or
more independent solutions which, according to the Ansatz, are expressible
as finite series in the derivatives of an elliptic Baker--Akhiezer
function, including an exponential factor.  (Cf.~(\ref{eq:jacobiHK}).)

In the context of~(\ref{eq:ellipticlame}), the elliptic-curve algebraic
form of the Lam\'e equation, this means that one hopes to be able to
construct a solution on the curve~$E_{g_2,g_3}$, except at a finite number
of values of the spectral parameter $B\in\mathbb{C}$, as a finite series in
the functions $\Phi^{(j)}\defeq (y\,\rd/\rd x)^j\Phi$, $j\ge0$, multiplied
by a factor $\exp\left[\kappa\int\rd x/y\right]$.  Here
$\Phi(\cdot,\cdot;x_0,y_0)$ is the fundamental multi-valued meromorphic
function on~$E_{g_2,g_3}$ introduced in~\S\,\ref{sec:algebraic}.  Actually,
a different but equivalent sort of series is easier to manipulate
symbolically.  By examination $\Phi^{(2)}=(2x+x_0) \Phi$, from which it
follows by induction on~$j$ that any finite series in $\Phi^{(j)}$,
$j\ge0$, is a combination (over polynomials in~$x$) of the basis functions
$\Phi,\Phi^{(1)}$.  This motivates the following definition.

\begin{definition}
A solution of the Lam\'e equation~(\ref{eq:ellipticlame}) on the elliptic
curve~$E_{g_2,g_3}$ is said to be an Hermite--Krichever solution if it is
of the form
\begin{equation}
\label{eq:HK}
\begin{split}
\left[\mathcal{A}(x)\Phi(x,y;x_0,y_0)+2\mathcal{B}(x)\Phi^{(1)}(x,y;x_0,y_0)\right]
&\exp\left[\kappa\int\frac{\rd x}y\right]\\
\qquad=
\left[\mathcal{A}(x)+\mathcal{B}(x)\left(\frac{y+y_0}{x-x_0}\right)\right]
\Phi(x,y;x_0,y_0)\,
&\exp\left[\kappa\int\frac{\rd x}y\right]
,
\end{split}
\end{equation}
for some $(x_0,y_0)\in E_{g_2,g_3}\setminus\{(\infty,\infty)\}$ and
$\kappa\in\mathbb{C}$.  Here $\mathcal{A},\mathcal{B}$ are polynomials.
\end{definition}

As defined, Hermite--Krichever solutions subsume most of the solutions
explored in \S\,\ref{sec:families}.  If $\kappa=0$, they reduce to
theta-twisted Lam\'e polynomials.  If $(x_0,y_0)=(e_\gamma,0)$ for
$\gamma=1,2,3$, in which case $\Phi,\Phi^{(1)}$ degenerate to
$\sqrt{x-e_\gamma},\allowbreak \frac12y/\sqrt{x-e_\gamma}$, they reduce to
twisted Lam\'e polynomials of Type~II\null.  If~both specializations are
applied, they reduce to ordinary Lam\'e polynomials of Type~II\null.  The
Lam\'e polynomials of Type~I, both ordinary and twisted, are not of the
Hermite--Krichever form, but they can be viewed as arising from a passage
to the $(x_0,y_0)\to(\infty,\infty)$ limit.

If $\mathcal{A}(x)=\sum_j a_jx^j$ and $\mathcal{B}(x)=\sum_j b_jx^j$,
substituting (\ref{eq:HK}) into~(\ref{eq:ellipticlame}) and equating the
coefficients of powers of~$x$ yields the coupled pair of recurrences
\begin{align}
\label{eq:HKrecurrence1}
\begin{split}
&(2j-\ell+1)(2j+\ell+2)\,a_j + [(4j+5)x_0 +\kappa^2 - B]\,a_{j+1} 
\\
&\qquad {}+ [-(j+\tfrac52)g_2 + 4x_0^2 - 2\kappa y_0](j+2)\,a_{j+2} - (j+2)(j+3)g_3
\,a_{j+3} \\
&\qquad {}+8\kappa(j+1)\,b_{j}    +4(\kappa x_0-y_0)(2j+3)\,b_{j+1}\\
&\qquad{}+2\left[ \kappa(4x_0^2-g_2) -2x_0y_0\right](j+2)\,b_{j+2}
=0,
\end{split}
\\[\jot]
\label{eq:HKrecurrence2}
\begin{split}
&(2j-\ell+2)(2j+\ell+3)\,b_j + [-(4j+7)x_0 +\kappa^2 - B]\,b_{j+1}  
\\
&\qquad {}+ [-(j+\tfrac32)g_2 - 4x_0^2 + 2\kappa y_0](j+2)\,b_{j+2} - (j+2)(j+3)g_3
\,b_{j+3}\\
&\qquad {}+ \kappa(2j+3)\,a_{j+1} -(2\kappa x_0-y_0)(j+2)\,a_{j+2}
=0.
\end{split}
\end{align}
If $\kappa=0$, (\ref{eq:HKrecurrence1})--(\ref{eq:HKrecurrence2}) reduce to
(\ref{eq:thetarecurrence1})--(\ref{eq:thetarecurrence2}), and if
$(x_0,y_0)=(e_\gamma,0)$, they reduce to
(\ref{eq:twistedrecurrence3})--(\ref{eq:twistedrecurrence4}).  If both
specializations are applied, they reduce to
(\ref{eq:recurrence3})--(\ref{eq:recurrence4}).

\begin{proposition}
For all $\ell\ge2$, Hermite--Krichever solutions can in~principle be
computed from these recurrences.  If $\ell$~is odd, resp.\ even, then
$\deg\mathcal{A},\mathcal{B}$ are ${(\ell-1)}/2,\allowbreak(\ell-3)/2$,
resp.\ $\ell/2-1,\ell/2-1$.  The coefficients are computed by setting the
coefficient of the highest power of~$x$ in $\mathcal{A}$,
resp.~$\mathcal{B}$, to unity, and working downward.
\end{proposition}

Unless $B,\kappa$ and the point $(x_0,y_0)$ are specially chosen, the
coefficients of negative powers of~$x$ may be nonzero.  But by examination,
they will be zero if the coefficients of~$x^{-1}$ in $\mathcal{A}$
and~$\mathcal{B}$ are both zero.  $a_{-1}=0$, $b_{-1}=0$ are equations
in~$B;\kappa;x_0,y_0$.  They are `compatibility conditions' similar to
those that appear in other applications of the Hermite--Krichever Ansatz.
Together with the identity $y_0^2=4x_0^3-g_2x_0-g_3$, they make~up a set of
three equations for these four unknowns.

Informally, one may eliminate any two of $B;\kappa;x_0,y_0$, and derive an
algebraic relation between the remaining two unknowns (involving $g_2,g_3$
of~course).  A~rigorous investigation must be more careful.  For~example,
if the ideal generated by the three equations contained a polynomial
involving only~$B$ (and~$g_2,g_3$), then a solution of the
Hermite--Krichever form would exist for very few values of~$B$
\mycite{Brezhnev2000}.  In~practice, this problem does not arise: except at a
finite number of values of~$B$ (at~most), the Ansatz can be
employed~\mycite{Gesztesy98}.  In~fact, in previous work an algebraic curve
in~$(B,\kappa)$ has been derived for each $\ell\le5$.  The
$(B,\kappa)$-curve is the one with the most physical significance, since
$B$~is a transformed energy and $\kappa$~is related to the crystal
momentum.  

Solving for $(x_0,y_0)$ as functions of~$(B,\kappa)$ reveals that $x_0$~is
a rational function of~$B$, and that if $\kappa$~is not identically zero,
$y_0$~is a rational function of~$B$, times~$\kappa$.  These facts can be
interpreted in~terms of the following seemingly different curve.

\begin{definition}
The $\ell$'th Lam\'e spectral curve $\Gamma_\ell\defeq\Gamma_\ell(g_2,g_3)$
is the hyperelliptic curve over $\mathbb{P}^1\ni B$ comprising all
$(B,\nu)$ satisfying $\nu^2=\mathsf{L}_\ell(B;g_2,g_3)$, where
$\mathsf{L}_\ell$~is the full Lam\'e spectral polynomial, of
degree~$2\ell+1$ in~$B$.  ($\Gamma_\ell$~was informally introduced
as~$\widetilde\Gamma_\ell$ in~\S\,\ref{sec:intro}, where the original
energy parameter~$E$ was used.)  $\Gamma_\ell$~will have genus~$\ell$
unless two roots of $\mathsf{L}_\ell(\cdot;g_2,g_3)$ coincide, i.e., unless
the Klein invariant~$J$ is a root of one of the two Cohn polynomials of
table~\ref{tab:cohn}, in which case the genus equals~$\ell-1$.
\end{definition}

For each~$\ell$, there must exist a parametrization of Lam\'e equation
solutions by a point~$(B,\nu)$ on the punctured curve
$\Gamma_\ell\setminus\{(\infty,\infty)\}$, by the general theory of Hill's
equation on~$\mathbb{R}$ \mycite{McKean79}.  For any finite-band
Schr\"odinger equation on an elliptic curve, including the integer-$\ell$
Lam\'e equation, the Baker--Akhiezer function~(\ref{eq:hh}) provides such a
parametrization of solutions.  In~the general theory, the parametrizing
hyperelliptic curve~$\Gamma$ for any finite-band Hill's equation arises
from {\em differential--difference bispectrality\/}: as a uniformization of
the relation between the energy parameter~$B$ and the crystal momentum~$k$
\mycite{Treibich2001orig}.  This curve~$\Gamma\ni(B,\nu)$ is defined by an
irrationality of the form $\nu^2={\sf L}(B)$, and $B$ and~$k$ are
meromorphic functions on~it; the former single-valued, and the latter
additively multi-valued.  The energy is computed from the degree-$2$ map
$B:\Gamma\to\mathbb{P}^1$ given by $(B,\nu)\mapsto B$, and the crystal
momentum from the formula
\begin{equation}
\label{eq:oint}
k= -\ri \oint\left[\frac12\, 
\left(
\frac{y+y_0(B,\nu)}{x-x_0(B,\nu)}
\right)
+ \kappa(B,\nu)
\right]\frac{dx}{y}
\Biggm/
\oint\frac{dx}{y}
,
\end{equation}
in which the line integrals on~$E_{g_2,g_3}$ are taken over the appropriate
fundamental loop.  Here $(B,\nu)\mapsto(x_0,y_0)$ is a certain projection
$\pi:\Gamma\to E_{g_2,g_3}$, and $\kappa:\Gamma\to\mathbb{P}^1$ is a
certain auxiliary meromorphic function.  These two morphisms of complex
manifolds are `odd' under the involution $(B,\nu)\mapsto(B,-\nu)$, i.e.,
$x_0(B,-\nu)=x_0(B,\nu)$ and $y_0(B,-\nu)=-y_0(B,\nu)$, and
$\kappa(B,-\nu)=-\kappa(B,\nu)$.  So $x_0$~must be a rational function
of~$B$, and each of $y_0$ and~$\kappa$ must be a rational function of~$B$,
times~$\nu$.

In the general theory of finite-band equations, the Baker--Akhiezer
uniformization is viewed as more fundamental than the Hermite--Krichever
Ansatz.  However, in the case of the integer-$\ell$ Lam\'e equation one can
immediately identify the curve in~$(B,\kappa)$, derived from the Ansatz as
explained above, with the $\ell$'th Lam\'e spectral curve
$\Gamma_\ell\ni(B,\nu)$.  It~is isomorphic to~it by a birational
equivalence of a simple kind: the ratio~$\kappa/\nu$ is a rational function
of~$B$.

This interpretation makes possible a geometrical understanding of each of
the types of Lam\'e spectral polynomial worked~out
in~\S\,\ref{sec:families}.  Due~to oddness, each of the finite Weierstrass
points $\{(B_s,0)\}_{s=0}^{2\ell}$ on~$\Gamma_\ell$, which correspond to
band edges, must be mapped by the projection $\pi_\ell:\Gamma_\ell\to
E_{g_2,g_3}$ to one of the finite Weierstrass points
$\{(e_\gamma,0)\}_{\gamma=1}^3$ or to~$(\infty,\infty)$.  Bearing in~mind
that Type~I Lam\'e polynomials, both ordinary and twisted, are not of the
Hermite--Krichever form, on~account of $(x_0,y_0)$ formally
equalling~$(\infty,\infty)$ and $\kappa$~equalling~$\infty$, one has the
following proposition.

\begin{proposition}
\mbox{}
\begin{enumerate}
\item The roots of the Type-{I} Lam\'e spectral polynomial $L_\ell^{\rm
I}(B)$ are the $B$-values of the finite Weierstrass points
$\{(B_s,0)\}_{s=0}^{2\ell}$ that are projected by ${\pi_\ell:\Gamma_\ell\to
E_{g_2,g_3}}$ to the infinite Weierstrass point~$(\infty,\infty)$.
Moreover, the roots of the Type-{I} twisted Lam\'e spectral polynomial
$Lt_\ell^{\rm I}(B)$ include the $B$-values of the finite non-Weierstrass
points that are projected to~$(\infty,\infty)$, and do not include the
$B$-value of any finite point that is not projected to~$(\infty,\infty)$.
\item For $\gamma=1,2,3$, the roots of the Type-{II} Lam\'e spectral
polynomial $L_\ell^{\rm II}(B;e_\gamma)$ are the $B$-values of the finite
Weierstrass points $\{(B_s,0)\}_{s=0}^{2\ell}$ that are projected by
${\pi_\ell:\Gamma_\ell\to E_{g_2,g_3}}$ to the finite Weierstrass
point~$(e_\gamma,0)$.  Moreover, the roots of the Type-{II} twisted Lam\'e
spectral polynomial $Lt_\ell^{\rm II}(B;e_\gamma)$ include the $B$-values
of the finite non-Weierstrass points that are projected to~$(e_\gamma,0)$,
and do not include the $B$-value of any finite point that is not projected
to~$(e_\gamma,0)$.
\item The roots of the theta-twisted Lam\'e spectral polynomial
$L\theta_\ell(B)$ include the $B$-values of the finite non-Weierstrass
points that are zeroes of\/ $\kappa_\ell:\Gamma_\ell\to\mathbb{P}^1$, and
do not include the $B$-value of any finite point that is not a zero.
\end{enumerate}
\label{prop:lastprop}
\end{proposition}
\begin{remark}
The phrasing of the proposition leaves open the possibilities that (1)~a
root of $Lt_\ell^{\rm I}(B)$ may be a root of~$L_\ell^{\rm I}(B)$, (2)~a
root of $Lt_\ell^{\rm II}(B,e_\gamma)$ may be a root of~$L_\ell^{\rm
II}(B,e_\gamma)$, and (3)~a root of~$L\theta_\ell^{\rm I}(B)$ may be the
$B$-value of a finite Weierstrass point, i.e., a band edge.  Generically
these three types of coincidence do not occur, but instances are not
difficult to find.  One is the case $\ell\equiv0\pmod3$ and $g_2=0$
(i.e.,~$J=0$), in which by examination $Lt^{\rm I}_\ell$ and~$L^{\rm
I}_\ell$ have the common root~${B=0}$.
\end{remark}

To proceed beyond the proposition, a significant result from finite-band
integration theory is needed.  {\em For the integer-$\ell$ Lam\'e equation,
the covering ${\pi_\ell:\Gamma_\ell\to E_{g_2,g_3}}$ and the auxiliary map
$\kappa_\ell:\Gamma_\ell\to\mathbb{P}^1$ are both of degree
$\ell(\ell+1)/2$, irrespective of the choice of elliptic
curve~$E_{g_2,g_3}$.}  From this fact, supplemented by
proposition~\ref{prop:lastprop}, the two propositions \ref{prop:twisted}
and~\ref{prop:thetatwisted}, which were left unproved
in~\S\,\ref{sec:families}, immediately follow.

The fibre over any point $(x_0,y_0)\in E_{g_2,g_3}$ must comprise
$\ell(\ell+1)/2$ points of~$\Gamma_\ell$, the counting being up~to
multiplicity.  Consider the fibre over $(\infty,\infty)$, which by
examination includes with unit multiplicity the point
$(B,\nu)=(\infty,\infty)$.  It~also includes each finite Weierstrass point
of~$\Gamma_\ell$ that corresponds to a Type-I Lam\'e polynomial.  As~was
shown in~\S\,\ref{sec:families}, the number of these up~to multiplicity,
$N_\ell^{\rm I}\defeq\deg L_\ell^{\rm I}$, equals $(\ell-1)/2$ if $\ell$~is
odd and $\ell/2+1$ if $\ell$~is even.  So the number of additional points
above $(\infty,\infty)$ is $\ell(\ell+1)/2-1-(\ell-1)/2=(\ell^2-1)/2$ if
$\ell$~is odd and $\ell(\ell+1)/2-1-(\ell/2+1)=\ell^2/2-2$ if $\ell$~is
even.  Since the projection~$\pi_\ell$ is odd under the involution
$(B,\nu)\mapsto(B,-\nu)$, these occur in pairs.  So $Nt_\ell^{\rm
I}\defeq\deg Lt_\ell^{\rm I}$ must equal $(\ell^2-1)/4$ if $\ell$~is odd
and $\ell^2/4-1$ if $\ell$~is even; which is the formula for~$Nt_\ell^{\rm
I}$ stated in proposition~\ref{prop:twisted}.  A~similar computation
applied to the fiber above any finite Weierstrass point $(e_\gamma,0)$
yields the formula for~$Nt_\ell^{\rm II}$ given in that proposition.

The formula for $N\theta_\ell\defeq\deg L\theta_\ell$ stated in
proposition~\ref{prop:thetatwisted} can also be derived with the aid of
proposition~\ref{prop:lastprop}.  Since the map
$\kappa:\Gamma_\ell\to\mathbb{P}^1$ has degree $\ell(\ell+1)/2$, the fibre
above~$0$ comprises that number of points, up~to multiplicity.  It~includes
each finite Weierstrass point of~$\Gamma_\ell$ that corresponds to a
Type-II Lam\'e polynomial (but not the Weierstrass points corresponding to
Type-I Lam\'e polynomials, since those are not of the Hermite--Krichever
form and formally have $\kappa=\infty$).  The number of these up~to
multiplicity is three times $N_\ell^{\rm II}\defeq\deg L_\ell^{\rm II}$,
which equals $3(\ell+1)/2$ if $\ell$~is odd and $3\ell/2$ if $\ell$~is
even.  So the number of additional points above~$0$ is
$\ell(\ell+1)/2-3(\ell+1)/2=(\ell+1)(\ell-3)/2$ if $\ell$~is odd and
$\ell(\ell+1)/2-3\ell/2=\ell(\ell-2)/2$ if $\ell$~is even.  They come in
pairs, and division by two yields for the formula for~$N\theta_\ell$ given
in proposition~\ref{prop:thetatwisted}.

A geometrized version of proposition~\ref{prop:cohnconj} can also be
proved, with the aid of an additional result that goes beyond the
Hermite--Krichever Ansatz.  {\em The Lam\'e spectral curve\/
$\Gamma_l(g_2,g_3)$ is nonsingular with genus~$\ell$ for generic values of
the Klein invariant $J=J(g_2,g_3)$, and when it degenerates to a singular
curve\/ $\Gamma^\circ_l\defeq\Gamma_\ell(g^\circ_2,g^\circ_3)$, the
singular curve has genus~$\ell-1$, with singularities that are limits as\/
$(g_2,g_3)\to(g_2^\circ,g_3^\circ)$ of Weierstrass points
of\/~$\Gamma_\ell(g_2,g_3)$.}  It~follows that the $\ell$'th Type-I and
Type-II Cohn polynomials, which characterize the pairs $(g_2,g_3)$ for
which the $\ell$'th Lam\'e operator on~$E_{g_2,g_3}$ has degenerate
algebraic spectrum of the specified type, i.e., for which
$\Gamma_\ell(g_2,g_3)$ has a pair of degenerate finite Weierstrass points
of the specified type, in~fact do more: they characterize the $(g_2,g_3)$
for which $\Gamma_\ell(g_2,g_3)$ is singular.  Due~to the reduction of the
genus by at~most unity, there can be, as proposition~\ref{prop:cohnconj}
states, no~multiple coincidences of the algebraic spectrum.

The problem of {\em explicitly\/} constructing an Hermite--Krichever
solution of the integer-$\ell$ Lam\'e equation, of the form~(\ref{eq:HK}),
will now be considered.  What are needed are the quantities
$x_0,y_0,\kappa$, or equivalently $x_0,y_0/\nu,\kappa/\nu$. Each of the
latter is a rational function of the spectral parameter~$B$.

One way of deriving these functions is to eliminate variables from the
system of three polynomial equations in $B;\kappa;x_0,y_0$, as explained
above.  Coupled with the spectral equation $\nu^2=\mathsf{L}_\ell(B)$, this
yields explicit expressions for the three desired functions.  By the
standards of polynomial elimination algorithms, this procedure is not
time-consuming.  It~is much more efficient than the manipulations of
compatibility conditions that other authors have employed.  Beginning with
\myciteasnoun{Halphen1886}, it~has been the universal practice to apply the
Hermite--Krichever Ansatz to the Weierstrassian form of the Lam\'e
equation, i.e., to~(\ref{eq:weierstrasslame}), rather than to the elliptic
curve form~(\ref{eq:ellipticlame}).  It~is now clear that this
Weierstrassian approach is far from optimal.  For~example, the computation
of the covering map $(B,\nu)\mapsto(x_0,y_0)$ in the case $\ell=5$, by
\myciteasnoun{Eilbeck94}, required seven hours of computer time.  The
just-sketched elliptic curve approach requires only a fraction of a second.

It turns~out that for constructing Hermite--Krichever solutions, this
revised elimination scheme is also not optimal.  Remarkably, at this point
{\em no~elimination needs to be performed at~all\/}, since the covering
$\pi_\ell:\Gamma_\ell\to E_{g_2,g_3}$ and auxiliary function
$\kappa_\ell:\Gamma_\ell\to\mathbb{P}^1$ can be computed directly from the
spectral polynomials of~\S\,\ref{sec:families}.

\begin{theoremL}
For all integer $\ell\ge1$, the covering map
$\pi_\ell:\Gamma_\ell(g_2,g_3)\to E_{g_2,g_3}$ appearing in the Ansatz maps
$(B,\nu)$ to~$(x_0,y_0)$ according to
\begin{align}
\label{eq:x0formula}
x_0(B;g_2,g_3)&=
e_\gamma + \frac4{[\ell(\ell+1)]^2}\,
\frac
{L_\ell^{\rm II}(B;e_\gamma,g_2,g_3)\, [Lt_\ell^{\rm II}(B;e_\gamma,g_2,g_3)]^2}
{L_\ell^{\rm I}(B;g_2,g_3)\, [Lt_\ell^{\rm I}(B;g_2,g_3)]^2}
\\
\label{eq:y0formula}
y_0(B,\nu;g_2,g_3)&=
\frac{16}{[\ell(\ell+1)]^3}
\left\{\frac{\prod_{\gamma=1}^3 Lt_\ell^{\rm II}(B;e_\gamma,g_2,g_3)}{[L_\ell^{\rm I}(B;g_2,g_3)]^2\, [Lt_\ell^{\rm I}(B;g_2,g_3)]^3}
\right\}\nu,
\end{align}
with $\gamma$ in\/ {\rm(\ref{eq:x0formula})} being any of\/ $1,2,3$.  The
auxiliary function\/~$\kappa_\ell:\Gamma_\ell\to\mathbb{P}^1$ is given by
\begin{equation}
\label{eq:kappaformula}
\kappa(B,\nu;g_2,g_3)= -\,
\frac{(\ell-1)(\ell+2)}{\ell(\ell+1)}
\left[
\frac{L\theta_\ell(B;g_2,g_3)}{L_\ell^{\rm I}(B;g_2,g_3)\,
Lt_\ell^{\rm I}(B;g_2,g_3)}
\right]\nu.
\end{equation}
\end{theoremL}

\begin{proof}
With the exception of the three $\ell$-dependent prefactors, such
as~$4/[\ell(\ell+1)]^2$, the formulas
(\ref{eq:x0formula})--(\ref{eq:kappaformula}) follow uniquely from
proposition~\ref{prop:lastprop}, regarded as a list of properties that
$\pi_\ell$ and~$\kappa_\ell$ must satisfy.  

$\pi_\ell$ must map each point $(B_s,0)$, where $B_s$~is a root
of~$L_\ell^{\rm I}(B)$, singly to~$(\infty,\infty)$, and each point
$(B_s,0)$, where $B_s$~is a root of~$L_\ell^{\rm II}(B;e_\gamma)$, singly
to~$(e_\gamma,0)$.  It~must also map each point $(B_t,\pm\nu_t)$, where
$B_t$~is a root of~$Lt_\ell^{\rm I}(B)$, singly to~$(\infty,\infty)$, and
each point $(B_t,\pm\nu_t)$, where $B_t$~is a root of~$Lt_\ell^{\rm
II}(B;e_\gamma)$, singly to~$(e_\gamma,0)$.  In~all these statements, the
counting is up~to multiplicity.

Also, $B\mapsto\kappa/\nu$ must map each point $(B',\pm\nu')$, where
$B'$~is a root of~$L\theta_\ell(B)$, singly to zero, and must map each
point $(B_s,0)$, where $B_s$~is a root of~$L_\ell^{\rm I}(B)$, and each
point $(B_t,\pm\nu_t)$, where $B_t$~is a root of~$Lt_\ell^{\rm I}(B)$,
singly to~$\infty$.  In~these statements as~well, the counting is up~to
multiplicity.

The $\ell$-dependent prefactors in
(\ref{eq:x0formula})--(\ref{eq:kappaformula}) can be deduced from the
leading-order asymptotic behavior of $x_0$ and~$\kappa/\nu$
as~$B\to\infty$.
\end{proof}

The remarkably simple formulas of the theorem permit Hermite--Krichever
solutions of the form~(\ref{eq:HK}) to be constructed for quite large
values of~$\ell$, since the Lam\'e spectral polynomials $L,Lt,L\theta$
(ordinary, twisted and theta-twisted) are relatively easy to work~out, as
\S\,\ref{sec:families} made clear.  Tables \ref{tab:spectral},
\ref{tab:twistedspectral} and~\ref{tab:thetatwistedspectral} may be
consulted.

It~should be stressed that in the formula~(\ref{eq:x0formula}) for~$x_0$,
the same right-hand side results, irrespective of which of the three values
of~$\gamma$ is chosen.  All terms explicitly involving~$e_\gamma$ will
cancel.  Of~course, all powers of~$e_\gamma$ higher than the second must
first be rewritten in~terms of $g_2,g_3$ by using the identity
$e_\gamma^3=\frac14({g_2e_\gamma+g_3})$.  In the same way, it is understood
that the numerator of the right-hand side of~(\ref{eq:y0formula}), the
terms of which are symmetric in $e_1,e_2,e_3$, should be rewritten in~terms
of~$g_2,g_3$.  (This can always be done: for~example, $e_1^2e_2^2 +
e_2^2e_3^2 + e_3^2e_1^2$ equals $g_2^2/16$.)

The application of Theorem~L to the cases $\ell=1,2,3$ may be illuminating.
\begin{itemize}
\item If $\ell=1$, then $(x_0,y_0)=(B,2\nu)$ and $\kappa=0$.  The map
$\pi_1:\Gamma_1\to E_{g_2,g_3}$ is a mere change of normalization, since
$\Gamma_1$ is isomorphic to~$E_{g_2,g_3}$; cf.~(\ref{eq:phisolns}).
\item If $\ell=2$, then
\begin{align}
\label{eq:x0l2case}
x_0 &= e_\gamma + \frac19\,\frac{(B+3e_\gamma)(B-6e_\gamma)^2}{B^2-3g_2}
= \frac{B^3+27g_3}{9(B^2-3g_2)},\\
\label{eq:y0l2case}
y_0 &=  \frac2{27}\,\frac{\prod_{\gamma=1}^3 (B-6e_\gamma)}{(B^2-3g_2)^2}\,\nu=
\frac{2(B^3-9 g_2 B-54 g_3)}{27 (B^2-3 g_2)^2}\,\nu,
\end{align}
and
$\kappa=-\left\{2\bigm/[3( B^2-3g_2)]\right\}\nu$.
\item If $\ell=3$, then
\begin{align}
\label{eq:x0l3case}
x_0 &= e_\gamma + \frac1{36}\,
\frac{(B^2-6e_\gamma  B+ 45e_\gamma^2 - 15g_2)(B^2-15e_\gamma  B-225e_\gamma^2 + \frac{75}4g_2)^2}{B (B^2-\frac{75}4g_2)^2}\nonumber\\[\jot]
&= \frac{{\dbinom{16 B^6+360 g_2 B^4+27000 g_3 B^3-3375 g_2^2
B^2\hskip0.1in\vrule width0in}{\hfill{}-303750 g_2 g_3  B-84375
g_2^3+2278125 g_3^2}}}{36\, B (4 B^2-75g_2)^2},\\[\jot]
y_0 &= \frac{1}{108}\,
\frac{\prod_{\gamma=1}^3(B^2-15e_\gamma  B-225e_\gamma^2 + \frac{75}4g_2)}{B^2 (B^2-\frac{75}4g_2)^3}\,\nu\nonumber\\[\jot]
\label{eq:y0l3case}
&= \frac{{\dbinom{16 B^6-1800 g_2 B^4-54000 g_3 B^3\hskip0.8in\vrule width0in}{\hfill{}-16875 g_2^2 B^2+421875 g_2^3-11390625 g_3^2}}}{27\, B^2 (4 B^2-75 g_2)^3}\,\nu,
\end{align}
and $\kappa= - \left\{10\bigm/[3B\,(4 B^2-75 g_2)]\right\}\nu$.
\end{itemize}
The formulas for $\ell=2,3$ were essentially known to Hermite.  Setting
$\ell=4,5$ in Theorem~L yields the less familiar and more complicated
formulas which \myciteasnoun{Enolskii94} and \myciteasnoun{Eilbeck94} derived
by eliminating variables from compatibility conditions.  Theorem~L readily
yields the covering map~$\pi_\ell$ and auxiliary function~$\kappa_\ell$ for
far larger~$\ell$.

\section{Hyperelliptic reductions}
\label{sec:reduction}

The cover $\pi_\ell:\Gamma_\ell(g_2,g_3)\to E_{g_2,g_3}$ introduced as part
of the Hermite--Krichever Ansatz, i.e., the map $(B,\nu)\mapsto(x_0,y_0)$,
is of independent interest, since explicit examples of coverings of
elliptic curves by higher-genus algebraic curves are few, and the problem
of determining which curves can cover~$E_{g_2,g_3}$, for either specified
or arbitrary values of the invariants~$g_2,g_3$, remains unsolved.
$\Gamma_\ell(g_2,g_3)$ generically has genus~$g=\ell$, as noted, and the
cover will always be of degree $N=\ell(\ell+1)/2$.  The formula for
$x_0=x_0(B;g_2,g_3)$ given in Theorem~L is consistent with this, since
$N$~equals $\max(N_\ell^{\rm I}+2Nt_\ell^{\rm I},\allowbreak N_\ell^{\rm
II}+2Nt_\ell^{\rm II})$, the maximum of the degrees in~$B$ of the numerator
and denominator of~$x_0$.  The degrees $N_\ell^{\rm I},N_\ell^{\rm II}$
were computed in~\S\,\ref{sec:families}, and the twisted degrees
$Nt_\ell^{\rm I},Nt_\ell^{\rm II}$ were also (see
proposition~\ref{prop:twisted}).

Since $\Gamma_\ell(g_2,g_3)$ is hyperelliptic (defined by the irrationality
$\nu^2 = \mathsf{L}_\ell(B;g_2,g_3)$) and $E_{g_2,g_3}$ is elliptic
(defined by the irrationality $y_0^2=4x_0^3-g_2x_0-g_3$), the
map~$\pi_\ell$ enables certain hyperelliptic integrals to be reduced to
elliptic ones.  Just as $E_{g_2,g_3}$ is equipped with the canonical
holomorphic $1$-form $\rd x_0/y_0$, so can $\Gamma_\ell(g_2,g_3)$ be
equipped with the holomorphic $1$-form $\rd B/\nu$.  Any integral of a
function in the function field of a hyperelliptic curve (here, any rational
function $R(B,\nu)$) against its canonical $1$-form is called a
hyperelliptic integral.  Hyperelliptic integrals are classified as follows
\mycite{Belokolos86orig}.  The linear space of meromorphic $1$-forms of the
form $R(B,\nu)\,\rd B/\nu$, i.e., of Abelian differentials, is generated by
$1$-forms of the first, second and third kinds.  These are (i)~{\em
holomorphic\/} $1$-forms, with no poles; (ii)~$1$-forms with one multiple pole;
and (iii)~$1$-forms with a pair of simple poles, the residues of which are
opposite in sign.  The indefinite integrals of (i)--(iii) are called
hyperelliptic integrals of the first, second and third kinds.  They
generalize the three kinds of elliptic integral
\mycite[chapter~17]{Abramowitz65}.

Hyperelliptic integrals of the first kind are the easiest to study, since
the linear space of holomorphic $1$-forms is finite-dimensional and is
spanned by $B^{r-1}\,\rd B/\nu$, $r=1,\dots,g$, where $g$~is the genus.
So there are only~$g$ independent integrals of the first kind.
A~consequence of the map $\pi_\ell:\Gamma_\ell(g_2,g_3)\to E_{g_2,g_3}$ is
that on any hyperelliptic curve of the form $\Gamma_\ell(g_2,g_3)$ there
are really only $g-1$ independent integrals of the first kind, modulo
elliptic integrals (considered trivial by comparison).  Changing variables
in $\int \rd x_0/y_0$, the elliptic integral of the first kind, yields
\begin{equation}
\label{eq:reduction}
\int\left[
\left(\frac{y_0}{\nu}\right)^{-1}
\frac{\rd x_0}{\rd B}
\right]\frac{\rd B}\nu
= \int\frac{\rd x_0}{y_0}.
\end{equation}
The quantity in square brackets is rational in~$B$, and in~fact is
guaranteed to be a polynomial in~$B$ of degree less than or equal to~$g-1$,
since the left-hand integrand is a pulled-back version of the right-hand
one, and must be a holomorphic $1$-form.  Equation~(\ref{eq:reduction}) is
a {\em linear constraint relation\/} on the $g$~basic hyperelliptic
integrals of the first kind. It~reduces the number of independent integrals
from $g$ to~$g-1$.

The cases $\ell=2,3$ of~(\ref{eq:reduction}) may be instructive.  The maps
$(B,\nu)\mapsto(x_0,y_0)$ were given in
(\ref{eq:x0l2case}),(\ref{eq:x0l3case}), and the degree-$(2\ell+1)$
spectral polynomials $\mathsf{L}_\ell(\cdot;g_2,g_3)$ follow from
table~\ref{tab:spectral}.  If $\ell=2$, one obtains the
hyperelliptic-to-elliptic reduction
\begin{equation}
\label{eq:l2}
\int \frac{[\tfrac32 B]\,\rd B}{\sqrt{(B^2-3g_2)(B^3-\frac94g_2B+\frac{27}4g_3)}}
=
\int \frac{\rd x_0}{\sqrt{4x_0^3 - g_2x_0  - g_3}},
\end{equation}
where the change of variables is performed by~(\ref{eq:x0l2case}).  If
$\ell=3$, one obtains
\begin{equation}
\label{eq:l3}
\int \frac{\left[3(B^2-\tfrac{15}4g_2)\right]\,\rd B}{\sqrt{\mathsf{L}_3(B;g_2,g_3)}}
=
\int \frac{\rd x_0}{\sqrt{4x_0^3 - g_2x_0  - g_3}},
\end{equation}
where the full spectral polynomial $\mathsf{L}_3(B;g_2,g_3)$ is
\begin{displaymath}
B\left(B^6-\tfrac{63}2 g_2 B^4+\tfrac{297}2 g_3 B^3 + \tfrac{4185}{16} g_2^2 B^2 -
\tfrac{18225}8 g_2 g_3 B - \tfrac{3375}{16}g_2^3 + \tfrac{91125}{16} g_3^2\right)
\end{displaymath}
and the change of variables is performed by~(\ref{eq:x0l3case}).  These
reductions were known to Hermite \mycite{Konigsberger1878,Belokolos86orig}.
More recently, the reductions induced by the $\ell=4,5$ coverings were
worked~out \mycite{Enolskii94,Eilbeck94}.  But the reductions with $\ell>5$
proved too difficult to compute.  Theorem~L makes possible the computation
of many such higher reductions.

The following proposition specifies the normalization of the pulled-back
$1$-form.  It~follows from the known leading-order asymptotic behavior of
$x_0,y_0/\nu$ as~$B\to\infty$.

\begin{table}
\caption{Polynomials specifying the holomorphic\/ $1$-forms pulled back from~$E_{g_2,g_3}$}
\hfill
\begin{tabular}{ll}
\hline
$\ell$ & $\hat P_\ell(B;g_2,g_3)$\\
\hline
$1$ & $1$\\
$2$ & $B$\\
$3$ & $B^2-\frac{15}4g_2$\\
$4$ & $B^3-\frac{91}4 g_2 B + \frac{175}2 g_3$\\
$5$ & $B^4-\frac{321}4 g_2 B^2 + \frac{2835}4 g_3 B + \frac{891}2 g_2^2$\\
$6$ & $B^5-\frac{861}4 g_2 B^3+\frac{12879}4 g_3 B^2+\frac{24255}4 g_2^2B-\frac{280665}4 g_2 g_3$\\
$7$ & $B^6-\frac{973}2 g_2 B^4+10813 g_3 B^3+\frac{681373}{16} g_2^2
B^2-\frac{2145143}{2} g_2 g_3 B+\frac{54071875}{16}
g_3^2-\frac{5417685}{16} g_2^3$\\
$8$ & $B^7-\frac{1953}2 g_2 B^5+29916 g_3 B^4+\frac{3335445}{16} g_2^2 B^3-\frac{34152435}4 g_2 g_3 B^2$\\
    & $\qquad{}+(-\frac{122490225}{16} g_2^3 + \frac{937038375}{16} g_3^2 )B+179425125 g_2^2 g_3$\\
\hline
\end{tabular}
\hfill
\label{tab:jacobian}
\end{table}

\begin{proposition}
For all integer $\ell\ge1$, the polynomial function
$P_\ell(B;g_2,g_3)\defeq[(y_0/\nu)^{-1}dx_0/dB](B;g_2,g_3)$ in the
hyperelliptic-to-elliptic reduction formula
\begin{displaymath}
\int \frac{P_\ell(B;g_2,g_3)\,\rd B}{\sqrt{\mathsf{L}_\ell(B;g_2,g_3)}}
=
\int \frac{\rd x_0}{\sqrt{4x_0^3 - g_2x_0  - g_3}},
\end{displaymath}
where the change of variables $x_0=x_0(B;g_2,g_3)$ is given by Theorem~L,
equals $\ell(\ell+1)/4$ times a polynomial $\hat P_\ell(B;g_2,g_3)$ which
is monic and of degree $\ell-1$ in~$B$.
\end{proposition}

The polynomials~$\hat P_\ell$ are listed in Table~\ref{tab:jacobian}.
$\hat P_4,\hat P_5$ agree with those found by Enol'skii et~al., if
allowance is made for a difference in normalization conventions.

A complete analysis of Lam\'e-derived elliptic covers will need to consider
exceptional cases of several kinds.  The covering curve
$\Gamma_\ell(g_2,g_3)$ generically has genus $g=\ell$, but if the Klein
invariant $J=g_2^3/(g_2^3-27g_3^2)$ is a root of one of the two Cohn
polynomials of table~\ref{tab:cohn}, the genus will be reduced to~$\ell-1$.
According to conjecture~\ref{conj:cohnconj}, this will happen, for
instance, if $\ell\equiv2\pmod3$ and $g_2=0$ (i.e., ${J=0}$), so that the
base curve $E_{g_2,g_3}$ is equianharmonic.  When $g$~is reduced
to~$\ell-1$ in this way, the linear space of holomorphic $1$-forms will be
spanned by $B^{r-1}(B-B_0)\,\rd B/\nu$, $r=1,\dotsc,\ell-1$, where $B_0$~is
the degenerate root of the spectral polynomial; but
(\ref{eq:reduction})~will still provide a linear constraint on the
associated hyperelliptic integrals.

Another sort of degeneracy takes place when the modular discriminant
$\Delta\defeq g_2^3-27g_3^2$ equals zero, i.e., when~$J=\infty$.  In~this
case $E_{g_2,g_3}$ will degenerate to a rational curve, due to two or more
of $e_1,e_2,e_3$ being coincident.  The Lam\'e-derived reduction formulas,
such as (\ref{eq:l2})--(\ref{eq:l3}), continue to apply.  (They are valid
though trivial even in the case $e_1=e_2=e_3$, in which $g_2=g_3=0$.)  So
these formulas include as special cases certain hyperelliptic-to-{\em
rational\/} reductions.

Subtle degeneracies of the covering map~$\pi_\ell$ can occur, even in the
generic case when $\Gamma_\ell$~has genus~$\ell$ and $E_{g_2,g_3}$ has
genus~$1$.  The {\em branching structure\/} of~$\pi_\ell$ is determined by
the polynomial~$\hat P_\ell$ of table~\ref{tab:jacobian}, which is
proportional to~$\rd x_0/\rd B$.  If $\hat P_\ell$ has distinct roots
$\{B^{(i)}\}_{i=1}^{\ell-1}$, then $\pi_\ell$~will normally have $2\ell-2$
simple critical points on~$\Gamma_\ell$, of the form
$\{(B^{(i)},\pm\nu^{(i)})\}_{i=1}^{\ell-1}$.  However, if any~$B^{(i)}$ is
located at a band edge, i.e., at a branch point of the hyperelliptic
$(B,\nu)$-curve, then $(B^{(i)},0)$ will be a double critical point.  This
appears to happen when $\ell\equiv0\pmod3$ and the base curve $E_{g_2,g_3}$
is equianharmonic; the double critical point being located
at~$(B,\nu)=(0,0)$.  Even if no root of~$\hat P_\ell$ is located at a band
edge, it is possible for it to have a double root, in which case each of a
pair of points $(B^{(i)},\pm\nu^{(i)})$ will be a double critical point.
By examination, this happens when $\ell=4$ and~$J=-2^25^4/3^553$.

A few hyperelliptic-to-elliptic reductions, similar to the quadratic
($N=2$) reduction of Legendre and Jacobi, can be found in handbooks of
elliptic integrals \mycite[\S\S\,575 and~576]{Byrd54}.  The Lam\'e-derived
reductions, indexed by~$\ell$, should certainly be included in any future
handbook.  It~is natural to wonder whether they can be generalized in some
straightforward way.  The problem of finding the genus-$2$ covers of an
elliptic curve was intensively studied in the nineteenth century, by
Weierstrass and Poincar\'e among many others, and one may reason by analogy
with results on $\ell=2$.  One expects that for all~$\ell\ge2$ and for
arbitrarily large~$N$, a~generic $E_{g_2,g_3}$ can be covered by some
genus-$\ell$ curve via a covering map of degree~$N$.  Each Lam\'e-derived
covering $\pi_\ell:\Gamma_\ell(g_2,g_3)\to E_{g_2,g_3}$ has
$N=\ell(\ell+1)/2$ and may be only a low-lying member of an infinite
sequence of coverings.  Generalizing the Lam\'e-derived coverings may be
possible even if one confines oneself to $N=\ell(\ell+1)/2$.  One can
of~course pre-compose with an automorphism of~$\Gamma_\ell(g_2,g_3)$ and
post-compose with an automorphism of~$E_{g_2,g_3}$ (a~modular
transformation).  But when $\ell=2$ a quite different covering map with the
same degree is known to exist \mycite{Belokolos86orig}.  $\pi_2$~has two
simple critical points on~$\Gamma_2$, but the other degree-$3$ covering map
has a single double critical point on its analogue of~$\Gamma_2$.  Both can
be generalized to include a free parameter
\mycite{Burnside1892,Belokolos2002orig}.  It~seems possible that when
$\ell>2$, similar alternatives to the Lam\'e-derived coverings may exist,
with degree $\ell(\ell+1)/2$ but different branching structures.

\section{Dispersion relations}
\label{sec:dispersion}

It is now possible to introduce dispersion relations, and determine the way
in which the Hermite--Krichever Ansatz reduces higher-$\ell$ to~$\ell=1$
dispersion relations.  The starting point is the fundamental multi-valued
function~$\Phi$ introduced in~\S\,\ref{sec:algebraic}.  As~noted, if the
parametrization point $(x_0,y_0)$ on the punctured elliptic curve
$E_{g_2,g_3}\setminus\{(\infty,\infty)\}$ is over $x_0=B\in\mathbb{C}$,
then $\Phi(\cdot,\cdot;x_0,y_0)$ will be a solution of the ${\ell=1}$ case
of the Lam\'e equation~(\ref{eq:ellipticlame}).  $E_{g_2,g_3}$ is defined
by $y^2 = 4x^3 - g_2x - g_3$, so the hypothesis here is that $(x_0,y_0)$
should equal $(B,\pm\sqrt{4B^3-g_2B-g_3})$.

In the Jacobi form (with independent variable~$\alpha$), resp.\ the
Weierstrassian form (with independent variable~$u$), the crystal
momentum~$k$ characterizes the behavior of a solution of the Lam\'e
equation under $\alpha\leftarrow\alpha+2K$, resp.\ $u\leftarrow u+2\omega$.
Both shifts correspond to motion around~$E_{g_2,g_3}$, along a fundamental
loop that passes between $(x,y)=(e_1,0)$ and~$(\infty,\infty)$, and cannot
be shrunk to a point.  (If~$e_1,e_2,e_3$ are defined by~(\ref{eq:e1e2e3}),
this will be because $y$~is positive on one-half of the loop, and negative
on the other.)  By definition, the solution will be multiplied by
$\xi\defeq \exp[\ri k(2K)]=\exp[\ri k(2\omega)]$.  It~follows from the
definition~(\ref{eq:Phi}) of~$\Phi$ that when~$\ell=1$,
\begin{equation}
\label{eq:floquetformula}
\xi = \exp[\ri k(2\omega)] = \exp\left[\frac12
\oint\left(\frac{y+y_0}{x-x_0}\right)\frac{\rd x}y \right].
\end{equation}
That is, when $\ell=1$ the crystal momentum is given by a complete elliptic
integral.  In the context of finite-band integration theory, this is a
special case of~(\ref{eq:oint}).

It was pointed~out in~\S\,\ref{sec:HK} that the spectral curve~$\Gamma_1$
that parametrizes $\ell=1$ solutions can be identified with $E_{g_2,g_3}$
itself, via the identification $(B,\nu)=(x_0,y_0/2)$.  This suggests a
subtle but important reinterpretation of~$k$.  In~\S\,\ref{sec:intro} it
was introduced as a function of the energy parameter, here~$B$, which is
determined only up~to integer multiples of~$\pi/K=\pi/\omega$, and which is
also undetermined as to sign.  If~the presence of $y_0=2\nu$
in~(\ref{eq:floquetformula}) is taken into account, it is clear that the
$\ell=1$ crystal momentum, called~$k_1$ henceforth, should be regarded as a
function not on~$\mathbb{P}^1\setminus\{\infty\}\ni B$, but rather
on~$\Gamma_1\setminus\{(\infty,\infty)\}\ni(B,\nu)$.  In this
interpretation, {\em the indeterminacy of sign disappears.}  The additive
indeterminacy, on~account of which $k_1$~is an elliptic function of the
second kind, remains but can be viewed as an artifact: it~is due to
$k_1\propto\log\xi$, where $\xi$~is the Floquet multiplier.  The behavior
of~$k_1$ near the puncture $(B,\nu)=(\infty,\infty)$ is easily determined.
It~follows from~(\ref{eq:floquetformula}) that as
$(B,\nu)\to(\infty,\infty)$, i.e., $(x_0,y_0)\to(\infty,\infty)$, each
branch of~$k_1$ is asymptotic to $-\ri\nu/B$ to leading order.  Since
$B=x_0,\allowbreak \nu=2y_0$ have double and triple poles there,
respectively, it follows that {\em each branch of~$k_1$ has a simple pole
at the puncture.}

The multiplier~$\xi$ is a true single-valued function on the punctured
spectral curve $\Gamma_1\setminus\{(\infty,\infty)\}$, and moreover is
entire.  One can write $\xi:\Gamma_1\setminus\{(\infty,\infty)\}\to
\mathbb{P}^1\setminus\{0,\infty\}$, since the multiplier is never zero.
Like~$k_1$, this function is not algebraic: it~necessarily has an essential
singularity at the puncture.  The $\left((B,\nu),\xi\right)$-curve over
$\Gamma_1\setminus\{(\infty,\infty)\}\ni(B,\nu)$, which is a single cover,
and the $(B,\xi)$-curve over~$\mathbb{C}\ni B$, which is a double cover,
are both {\em transcendental curves\/}.

The crystal momentum for each integer $\ell\ge1$ may similarly be viewed as
an additively multi-valued function on the punctured spectral curve
$\Gamma_\ell\setminus\{(\infty,\infty)\}$.  It~will be written as
$k_\ell(B,\nu;g_2,g_3)$, with the understanding that for this to be
well-defined, $B,\nu$~must be related by the spectral curve relation
$\nu^2=\mathsf{L}_\ell(B;g_2,g_3)$.  The quantity $k_\ell(B,\nu;g_2,g_3)$
will not be undetermined as to sign.  Suppose now that the projections
${\pi_\ell:\Gamma_\ell\to E_{g_2,g_3}}$ of the Hermite--Krichever Ansatz
are regarded as maps ${\pi_\ell:\Gamma_\ell\to\Gamma_1}$. That~is,
$\pi_\ell$~maps $(B,\nu)\in\Gamma_\ell$ to the point
$(B',\nu')\defeq(x_0,y_0/2)\in\Gamma_1$.  The reductions
$(B,\nu)\mapsto(B',\nu')$ for $\ell=2,3$, for~example, follow from
(\ref{eq:x0l2case})--(\ref{eq:y0l3case}).

\begin{proposition}
\label{prop:prefinal}
If the integration of the Lam\'e equation on the elliptic
curve~$E_{g_2,g_3}$, for integer $\ell\ge1$, can be accomplished in the
framework of the Hermite--Krichever Ansatz by maps $\pi_\ell:\Gamma_\ell\to
\Gamma_1$ and $\kappa_\ell:\Gamma_\ell\to \mathbb{P}^1$, where $\pi_\ell$
and~$\kappa_\ell$ map the point $(B,\nu)$ to
$\left(B_\ell(B;g_2,g_3),\allowbreak \nu_\ell(B,\nu;g_2,g_3)\right)$ and
$\hat\kappa_\ell(B;g_2,g_3)\nu$, respectively, then the dispersion relation
for the Hermite--Krichever solutions will be $k=k_\ell(B,\nu;g_2,g_3)$, in
which $k_\ell$ can be expressed in~terms of $k_1$ by
\begin{equation}
\label{eq:prekldef}
k_\ell(B,\nu;g_2,g_3) = k_1\!\left(B_\ell(B,\nu;g_2,g_3),\nu_\ell(B,\nu;g_2,g_3)\right) -\ri \hat\kappa_\ell(B,\nu;g_2,g_3)\nu.
\end{equation}
\end{proposition}

This proposition follows immediately from the form of the
Hermite--Krichever solutions~(\ref{eq:HK}).  The first term
in~(\ref{eq:prekldef}) arises from the $\Phi,\Phi'$ factors, and the second
from the exponential.  The factors ${\cal A},{\cal B}$ in~(\ref{eq:HK}) do
not contribute to the crystal momentum.  Equation~(\ref{eq:prekldef}) could
also be derived from the general theory of finite-band integration,
specifically from the formula~(\ref{eq:oint}).  However, a derivation from
the Hermite--Krichever Ansatz seems more natural in the present context.

The effort expended in replacing two-valuedness by single-valuedness is
justified by the following observation.  As~a function of~$B$ alone, rather
than of the pair~$(B,\nu)$, each of the two terms of~(\ref{eq:prekldef})
would be undetermined as to sign.  This ambiguity could cause confusion and
errors.  The present formulation, though a bit pedantic, facilitates the
determination of the correct relative sign.

The form of the $(B,\nu)\mapsto(B',\nu')$ map assumed in the proposition is
of~course the form supplied by Theorem~L\null.  Substituting the formulas
of the theorem into~(\ref{eq:prekldef}) yields an explicit expression
for~$k_\ell$.  It follows readily from this expression that for all
integer~$\ell\ge1$, each branch of~$k_\ell$ on the hyperelliptic
$(B,\nu)$-curve~$\Gamma_\ell(g_2,g_3)$ satisfies
$k_\ell\sim-\ri\nu/B^\ell$, $(B,\nu)\to(\infty,\infty)$.  Since $B,\nu$
have double and order-$(2\ell+1)$ poles at the puncture $(\infty,\infty)$,
respectively, this implies that irrespective of~$\ell$, each branch of the
crystal momentum has a simple pole at the puncture.

\section{Band structure of the Jacobi form}
\label{sec:jacobi}

The results of \S\S\,\ref{sec:families} through \ref{sec:dispersion} were
framed in~terms of the elliptic-curve algebraic form of the Lam\'e
equation.  Most work on Lam\'e dispersion relations has used the Jacobi
form instead, and has led accordingly to expressions involving Jacobi theta
functions.  To~derive dispersion relations that can be compared with
previous work, the formulation of~\S\,\ref{sec:dispersion} must be
converted to the language of the Jacobi form.

The relationships among the several forms were sketched
in~\S\,\ref{sec:algebraic}.  In the Weierstrassian and Jacobi forms, the
Lam\'e equation is an equation on~$\mathbb{C}$ with doubly periodic
coefficients, rather than an equation on the curve~$E_{g_2,g_3}$.  In~the
conversion to the Jacobi form the invariants $g_2,g_3$ are expressed
in~terms of the modular parameter~$m$ by~(\ref{eq:g2g3}), with $A=1$ by
convention.  The coordinate~$x$ on~$E_{g_2,g_3}$ is interpreted as the
function~$\wp(u;g_2,g_3)$, i.e., as $m\,{\rm sn}^2(\alpha|m)-\frac13(m+1)$,
where $u\in\mathbb{C}$ and $\alpha\defeq u-\ri\mathsf{K}'(m)\in\mathbb{C}$
are the respective independent variables of the Weierstrassian and Jacobi
forms.  The holomorphic differential $\rd x/y$ corresponds to $\rd u$
or~$\rd\alpha$, and the derivative $y\,\rd/\rd x$ to $\rd/\rd u$
or~$\rd/\rd\alpha$.  The coordinate~$y=(y\,\rd/\rd x)x$ on~$E_{g_2,g_3}$ is
interpreted as~$\wp'(u;g_2,g_3)$, i.e., as the doubly periodic function
$2m\,{\rm sn}(\alpha|m)\,{\rm cn}(\alpha|m)\,{\rm dn}(\alpha|m)$ on the
complex $\alpha$-plane.  The functions $\sqrt{x-e_\gamma}$, $\gamma=1,2,3$,
correspond to $-\ri\,{\rm dn}(\alpha|m)$, $-\ri\,m^{1/2}{\rm cn}(\alpha|m)$
and~$m^{1/2}\,{\rm sn}(\alpha|m)$.
\begin{equation}
\label{eq:BE2}
B=-E+\tfrac13\ell(\ell+1)(m+1)
\end{equation}
relates the accessory parameters $B,E$ of the different forms.

With these formulas, it is easy to convert the Lam\'e polynomials of
table~\ref{tab:polys} to polynomials in ${\rm sn}(\alpha|m)\allowbreak,{\rm
cn}(\alpha|m)\allowbreak,{\rm dn}(\alpha|m)$, and the spectral polynomials
of table~\ref{tab:spectral} to polynomials in~$E$, for comparison with the
list given by \myciteasnoun[\S\,9.3.2]{Arscott64}.

\begin{definition}
The Jacobi-form spectral polynomial $\widetilde{\mathsf{L}}_\ell(E|m)$ is
the {\em negative\/} of the spectral polynomial
$\mathsf{L}_\ell(B;g_2,g_3)$, when $B;g_2,g_3$ are expressed in terms
of~$E,m$.  It~is a monic degree-$(2\ell+1)$ polynomial in~$E$ with
coefficients polynomial in~$m$, and can be regarded as
$\prod_{s=0}^{2\ell}[E-E_s(m)]$, where the roots $\{E_s\}$ are the values
of the energy~$E$ for which there exists a Lam\'e polynomial solution of
the Lam\'e equation, counted with multiplicity.  (The negation is due~to
the relative minus sign in the $B\leftrightarrow E$
correspondence~(\ref{eq:BE2}).)
\end{definition}

\begin{definition}
The $\ell$'th Jacobi-form spectral curve
$\widetilde\Gamma_\ell\defeq\widetilde\Gamma_\ell(m)$ is the hyperelliptic
curve over $\mathbb{P}^1\ni E$ comprising all $(E,\tilde\nu)$ satisfying
$\tilde\nu^2=\widetilde{\mathsf{L}}_\ell(E|m)$.  In~the usual case when
$m$~is real, $\tilde\nu$~will be real if $E$~is in a band, and non-real if
$E$~is in a lacuna.  In~both cases it is determined only up~to negation.
By convention, the correspondence between the curve
$\widetilde\Gamma_\ell\ni(E,\tilde \nu)$ and the previously introduced
curve $\Gamma_\ell\ni(B,\nu)$, which was defined by
$\nu^2=\mathsf{L}_\ell(B;g_2,g_3)$, is given by $\nu=+\ri\tilde\nu$.
\end{definition}

The following cases are examples.  When $\ell=1,2,3$, the spectral
polynomial factors over the integers into polynomials at~most quadratic
in~$E$.  In~full,
\begin{align}
\label{eq:specl1}
\widetilde{\mathsf{L}}_1(E|m) &= 
(E-1)(E-m)(E-m-1)\\
\label{eq:specl2}
\widetilde{\mathsf{L}}_2(E|m) &= 
\left[E^2-4(m+1)E+12m\right](E-m-1)(E-4m-1)(E-m-4)
\\
\label{eq:specl3}
\widetilde{\mathsf{L}}_3(E|m) &= 
(E-4m-4)\left[E^2-2(2m+5)E+3(8m+3)\right]\\
&\qquad{}\times\left[E^2-2(5m+2)E+3(3m^2+8m)\right]\nonumber\\
&\qquad{}\times\left[E^2-10(m+1)E+3(3m^2+26m+3)\right].\nonumber
\end{align}
In (\ref{eq:specl2}) and~(\ref{eq:specl3}) the first factor arises from
$L^{\rm I}_\ell(B;g_2,g_3)$ and the remaining three from the factors
$L^{\rm II}_\ell(B;e_\gamma,g_2,g_3)$, $\gamma=1,2,3$.
In~(\ref{eq:specl1}) there is no Type~I factor.  The polynomials
(\ref{eq:specl1})--(\ref{eq:specl3}) agree with those obtained by Arscott.

The derivation of the Jacobi-form spectral polynomial
$\widetilde{\mathsf{L}}_\ell(E|m)$ from~$\mathsf{L}_\ell(B;g_2,g_3)$ sheds
light on a regularity noticed by \myciteasnoun[\S\,7]{Ince40a}, which
arises in the lemniscatic case~$m=\frac12$.  Ince observed that
if~$\ell\le6$, at~least, then $\widetilde{\mathsf{L}}_\ell(E|\frac12)$ has
an integer root, namely $E=\ell(\ell+1)/2$.  In~fact, this is the case for
all integer~$\ell$.  By~(\ref{eq:BE2}), the presence of this root is
equivalent to the full spectral polynomial $\mathsf{L}_\ell(B;g_2,0)$
having $B=0$ as a root.  But if~$m=\frac12$, it follows
from~(\ref{eq:e1e2e3}) that~$e_2=0$.  A~glance at the pattern of
coefficients in table~\ref{tab:spectral} reveals that if $g_3=0$ and a
singular point~$e_\gamma$ also equals zero, then either the Type-I spectral
polynomial~$L_\ell^{\rm I}(B;g_2,g_3)$ (if~$\ell\equiv0,3\mod4$) or one of
the three Type-II spectral polynomials~$L_\ell^{\rm
II}(B;e_\gamma,g_2,g_3)$ (if~$\ell\equiv1,2\pmod4$) will necessarily have
$B=0$ as a root.

Dispersion relations in their Jacobi form can now be investigated.  Recall
that if $\ell=1$, the Jacobi-form Lam\'e equation~(\ref{eq:jacobilame}) has
$\widetilde\Phi(\cdot;\alpha_0|m)$ as a solution, where the theta
quotient~$\widetilde\Phi$ (the Jacobi-form version of Halphen's {\em
l'\'el\'ement simple\/}) is defined in~(\ref{eq:jacobiPhi}), and the
multi-valued parameter~$\alpha_0$ is defined by ${\rm
dn}^2(\alpha_0|m)=E-m$.  This solution has crystal momentum~$k=k_1$ equal
to $-\ri\,{\rm Z}(\alpha_0|m)+\pi/2\mathsf{K}(m)$, which is undetermined
up~to sign, and is also determined only up~to integer multiples
of~$\pi/\mathsf{K}(m)$.  The sign indeterminacy is due to ${\rm
dn}^2(\cdot|m)$ being even.  This causes $\alpha_0$ to be undetermined
up~to sign, and $k_1$ as~well, since the function ${\rm Z}(\cdot|m)$ is
odd.

The parametrization point~$\alpha_0$, or the equivalent point
$u_0\defeq\alpha_0+\ri{\mathsf{K}}'(m)$ of the Weierstrassian form,
corresponds to the parametrization point $(x_0,y_0)$ of the fundamental
function~$\Phi$ on the elliptic curve~$E_{g_2,g_3}$.  The correspondence is
the usual one: $x_0=\wp(u_0;g_2,g_3)$, $y_0=\wp'(u_0,g_2,g_3)$.  The first
of these two equations says that $x_0=m\,{\rm
sn}^2(\alpha_0|m)-\frac13(m+1)$, and the latter that $y_0=2m\,{\rm
sn}(\alpha_0|m)\,{\rm cn}(\alpha_0|m)\,{\rm dn}(\alpha_0|m)$.  The formula
which computes $\alpha_0$ from~$E$, namely ${\rm dn}^2(\alpha_0|m)=E-m$, is
readily seen to be a translation to the Jacobi form of the familiar
condition $x_0=B$, which simply says that the parametrization point
$(x_0,y_0)$ must be `over' $B\in\mathbb{C}$.

The correspondence between the Jacobi and elliptic-curve forms motivates
the following reinterpretation of the crystal momentum of the fundamental
solution~$\widetilde\Phi$, which is modelled on the reinterpretation of the
last section.  $k_1$~should be viewed as a function not of the energy
$E\in\mathbb{C}$, but rather of a point $(E,\tilde\nu)$ on the punctured
Jacobi-form spectral curve
$\widetilde\Gamma_1(m)\setminus\{(\infty,\infty)\}$.  There are two such
points for each energy~$E$, except when $E$~is a band edge.  This is the
source of the sign ambiguity in the parameter~$\alpha_0$.  Since
$y_0=2\nu=2\ri\tilde\nu$, the equation
\begin{equation}
m\,{\rm sn}(\alpha_0|m)\,{\rm cn}(\alpha_0|m)\,{\rm dn}(\alpha_0|m) =
\ri\tilde\nu
\end{equation}
determines a unique sign for~$\alpha_0$, provided that $\tilde\nu$ is
specified in addition to~$E$.  $k_1$~will be written as
$k_1(E,\tilde\nu|m)$, with the understanding that for this to be
well-defined, the pair $E,\tilde\nu$~must be related by the spectral curve
relation $\tilde\nu^2=\widetilde{\mathsf{L}}_\ell(E|m)$.  The additively
undetermined quantity $k_1(E,\tilde\nu|m)$ will not be undetermined as to
sign.  It~is easily checked that on each branch, $k_1\sim\tilde\nu/E$ as
$(E,\tilde\nu)\to(\infty,\infty)$.

\begin{definition}
A solution of the Jacobi-form Lam\'e equation~(\ref{eq:jacobilame}) is said
to be an Hermite--Krichever solution if it is of the form
\begin{equation}
\label{eq:HKmodlast}
\left[\widetilde{\cal A}\left({\rm sn}^2(\alpha|m)\right)
\widetilde\Phi(\alpha;\alpha_0|m) + 2\widetilde{\cal B}\left({\rm
sn}^2(\alpha|m)\right)
\widetilde\Phi'(\alpha;\alpha_0|m)\right]\exp(\kappa\,\alpha),
\end{equation}
for some $\alpha_0\in\mathbb{C}$ and~$\kappa\in\mathbb{C}$.  Here
$\widetilde{\mathcal{A}},\widetilde{\mathcal{B}}$ are polynomials, and
$\widetilde\Phi'\defeq (\rd/\rd\alpha)\widetilde\Phi$.
\end{definition}

The expression~(\ref{eq:HKmodlast}) is a replacement for the original
Jacobi-form expression (\ref{eq:jacobiHK}), to which it is equivalent.
Regardless of which is used, it is easy to compute the crystal momentum of
an Hermite--Krichever solution.  The momentum computed
from~(\ref{eq:HKmodlast}) will be $[-\ri\,{\rm
Z}(\alpha_0|m)+\pi/2\mathsf{K}(m)]-\ri\kappa$, up~to additive
multi-valuedness.  The first term arises from the
$\widetilde\Phi,\widetilde\Phi'$ factors, and the second from the
exponential.  The factors $\widetilde{\cal A},\widetilde{\cal B}$ do not
contribute, since ${\rm sn}^2(\alpha|m)$ is periodic in~$\alpha$ with
period~$2\mathsf{K}(m)$.

The Jacobi form of the Hermite--Krichever Ansatz asserts that for all
integer~$\ell$ and $m\in\mathbb{C}\setminus\{0,1\}$, there is an
Hermite--Krichever solution for all but a finite number of values of the
energy~$E$.  On~the elliptic curve~$E_{g_2,g_3}$, these solutions were
constructed from two maps: a~projection $\pi_\ell:\Gamma_\ell\to
E_{g_2,g_3}$ and an auxiliary function $\kappa_\ell:\Gamma_\ell\to
\mathbb{P}^1$.  But $\pi_\ell$~should really be regarded as a map from
$\Gamma_\ell$ to~$\Gamma_1$, on~account of the correspondence between
$E_{g_2,g_3}$ and~$\Gamma_1$ provided by $(x_0,y_0)=(B,2\nu)$.  The
following is the Jacobi-form version of proposition~\ref{prop:prefinal}.

\begin{proposition}
\label{prop:final}
Suppose that the integration of the Lam\'e equation on the elliptic
curve~$E_{g_2,g_3}$, for integer $\ell\ge1$, can be accomplished in the
framework of the Hermite--Krichever Ansatz by the maps
$\pi_\ell:\Gamma_\ell\to \Gamma_1$ and $\kappa_\ell:\Gamma_\ell\to
\mathbb{P}^1$, where $\pi_\ell$ and~$\kappa_\ell$ map the point $(B,\nu)$
to $\left(B_\ell(B;g_2,g_3),\allowbreak \nu_\ell(B,\nu;g_2,g_3)\right)$ and
$\hat\kappa_\ell(B;g_2,g_3)\nu$, respectively.  Then the dispersion
relation for the solutions of the Jacobi form of the Lam\'e equation will
be $k=k_\ell(E,\tilde\nu|m)$, where up~to additive multi-valuedness
\begin{equation}
\label{eq:kldef}
k_\ell(E,\tilde\nu|m) = k_1\bigl({\cal E}_\ell(E|m),\tilde\nu_\ell(E,\tilde\nu|m)\!\bigm|\!m\bigr) +
\hat\kappa_\ell(E|m)\tilde\nu,
\end{equation}
in which
\begin{align}
\label{eq:newE}
{\cal E}_\ell(E|m)&\defeq
-B_\ell\!\left(-E+\tfrac13\ell(\ell+1)(m+1);g_2(m),g_3(m)\right) + \tfrac23(m+1),\\
\tilde\nu_\ell(E,\tilde\nu|m)&\defeq-\ri\nu_\ell\left(-E+\tfrac13\ell(\ell+1)(m+1), \ri\tilde\nu; g_2(m),g_3(m)\right),\\
\hat\kappa_\ell(E|m)&\defeq\hat\kappa_\ell\!\left(-E+\tfrac13\ell(\ell+1)(m+1); g_2(m),g_3(m)\right).
\end{align}
\end{proposition}

%In~the common case $E\in\mathbb{R}$, $k\in\mathbb{R}$, $\alpha_0$~is
%typically non-real \mycite[figure~5]{Li00}.

The formula~(\ref{eq:kldef}) follows by inspection.  The
projection~$\pi_\ell$ reduces the integration of the Lam\'e equation to the
integration of an $\ell=1$ equation, the `$B$'~parameter of which equals
$B_\ell(B;g_2,g_3)$.  By~(\ref{eq:BE2}), the `$E$'~parameter of the
$\ell=1$ equation will be the right-hand side of~(\ref{eq:newE}).  The two
terms of~(\ref{eq:kldef}) are simply the two terms of $[-\ri\,{\rm
Z}(\alpha_0|m)+\pi/2\mathsf{K}(m)]-\ri\kappa$.  The equality
$\nu=\ri\tilde\nu$ has been used.

It is straightforward to apply proposition~\ref{prop:final} to the cases
$\ell=2,3$, since the coverings $\pi_2,\pi_3$ and auxiliary functions
$\kappa_2,\kappa_3$ were worked~out in~\S\,\ref{sec:HK}.  A~brief
discussion of the $\ell=2$ case should suffice.  After some algebra, one
finds
\begin{align}
{\cal E}_2(E|m) &= \frac {E^3-12(m+1)^2E - 4(m+1)(4m^2 - 19m +4)} {9[E^2 - 4(m+1)E + 12m]},\\
\tilde\nu_2(E,\tilde\nu|m) &= 
-\,\frac{(E + 2 m - 4)(E - 4 m + 2)(E - 4 m - 4)}{27\left[E^2  - 4(  m + 1) E + 12 m\right]^2}
\,\tilde\nu,\\
\hat\kappa_2(E|m) &= 
-\,\frac{2}{3[E^2 - 4(m+1)E + 12m]},
\end{align}
from which $k_2(E,\tilde\nu|m)$ may be computed by~(\ref{eq:kldef}).
Like~$k_1$, $k_2$~is determined only up~to integer multiples of
$\pi/K\defeq\pi/{\mathsf K}(m)$.  Each branch of~$k_2$ has the property
that $k_2(E,\tilde\nu|m)\sim\pm E^{1/2}$, ${E}\to+\infty$, with `$\pm$'
determined by the sign of~$\tilde\nu=\tilde\nu({E})$.  This is a special
case of a general fact: for all integer $\ell\ge1$,
$k_\ell(E,\tilde\nu|m)\sim\pm E^{1/2}$, $E\to+\infty$, since each branch
of~$k_\ell$ is asymptotic to $(-)^{\ell-1}\tilde\nu/E^\ell$
as~$(E,\tilde\nu)\to(\infty,\infty)$.

\begin{figure}
\centerline{\epsfig{file=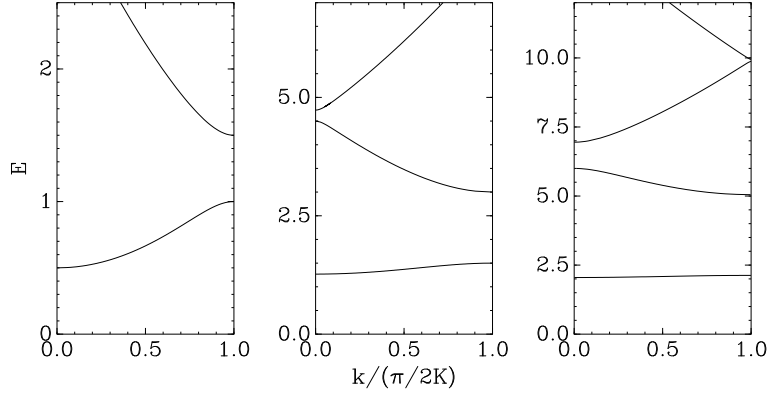,height=2.1in}}
\caption{Dispersion relations for $\ell=1,2,3$ in the lemniscatic case~$m=\tfrac12$.}
\label{fig:only}
\end{figure}

The real portions of the dispersion relations $k=k_1(E,\tilde\nu|\frac12)$,
$k_2(E,\tilde\nu|\frac12)$ and $k_3(E,\tilde\nu|\frac12)$ are graphed in
figure~\ref{fig:only}.  For ease of viewing, each crystal momentum is
regarded as lying in the interval~$[0,\pi/2K]$; which is equivalent to
choosing the sign of~$\tilde\nu=\tilde\nu(E)$ in an $E$-dependent way.  As
(\ref{eq:specl1})--(\ref{eq:specl3}) imply, the two $\ell=1$ bands are
$[\frac12,1],\allowbreak[\frac32,\infty)$, the three $\ell=2$ bands are
$[3-\sqrt3,\frac32],\allowbreak[3,\frac92],\allowbreak[3+\sqrt3,\infty)$,
and the four $\ell=3$ bands are $[\frac92 - \sqrt6,
6-\sqrt{15}],\allowbreak [\frac{15}2-\sqrt6, 6],\allowbreak
[\frac{9}2+\sqrt6, 6+\sqrt{15}],\allowbreak [\frac{15}2+\sqrt6, \infty)$.

%[2.05, 2.13]  [5.05, 6]  [6.95, 9.87]  [9.95, infty)

The $\ell=1$ graph agrees with that of \myciteasnoun[figure~6]{Li00}, and for
confirmation, with that of \myciteasnoun[figure~1]{Sutherland73}.
Unfortunately, the $\ell=2$ graph disagrees with that of Li et~al.\ in the
placement or direction of curvature of each of the two upper bands.  The
algorithm they used for reducing $\ell=2$ to~$\ell=1$, which was based on
Hermite's solution of the Jacobi-form Lam\'e equation
\mycite[\S\,23.71]{Whittaker27}, evidently yielded incorrect results for
these bands.  It~appears that for the middle band, at~least, the
discrepancy can be traced to an incorrect choice of relative sign for the
two terms of~$k=k_2$.  The reinterpretation of the crystal momentum as a
function on the spectral curve, rather than a function of the energy,
eliminates such sign ambiguities.

\section{Summary and final remarks}
\label{sec:conclusions}

A new approach to the closed-form solution of the Lam\'e equation has been
introduced.  Theorem~L provides a formula for the covering map of the
Hermite--Krichever Ansatz in~terms of certain polynomials which are of
independent interest, namely twisted spectral polynomials.  The theorem
permits an efficient computation of Lam\'e dispersion relations, of a mixed
symbolic--numerical kind.  Cohn polynomials, which are a new concept, have
also been introduced.  The roots of such a polynomial are the points in
elliptic moduli space at~which a Lam\'e spectral polynomial has a double
root, so that the Lam\'e spectral curve becomes singular.  Twisted and
theta-twisted Cohn polynomials could be defined, as~well.

The approach of this paper can be extended from the Lam\'e equation to the
Heun equation, which as a differential equation on the elliptic
curve~$E_{g_2,g_3}$ has up~to four regular singular points, positioned at
the finite Weierstrass points $\{(e_\gamma,0)\}_{\gamma=1}^3$ as~well as
at~$(\infty,\infty)$.  Its Weierstrassian form is called the
Treibich--Verdier equation \mycite{Smirnov2001}, and its Jacobi form,
at~least when only two of the Weierstrass points are singular points, the
associated Lam\'e equation \mycite[\S\,7.3]{Magnus79}.

The `four triangular numbers' condition for the Heun equation to have the
finite-band property, due to \myciteasnoun{Treibich92} and
\myciteasnoun{Gesztesy95a}, is now well known.  In~the finite-band case,
the number of points in the algebraic spectrum has been computed up~to
multiplicity \mycite{Gesztesy95a}.  The corresponding band-edge solutions
are Heun polynomials.  Applying the Hermite--Krichever Ansatz to the
finite-band Heun equation leads to a greater variety of coverings
of~$E_{g_2,g_3}$ than arise in the solution of the integer-$\ell$ Lam\'e
equation; for~example, coverings by a genus-$2$ hyperelliptic curve that
have degrees~$3,4,5$ \mycite{Belokolos2002orig}.  These coverings play a
role in the construction of elliptic soliton solutions of certain nonlinear
evolution equations that occur in fibre optics \mycite{Christiansen2000}.
A~treatment of the Heun equation along the lines of this paper will appear
elsewhere.

\begin{acknowledgements}
This work was partially supported by NSF grant PHY-0099484.  The symbolic
computations were performed with the aid of the {\sc Macsyma} computer
algebra system.  The helpful comments of an anonymous referee are
gratefully acknowledged.
\end{acknowledgements}

%\bibliographystyle{roysoc}
%\bibliography{general}

% NOTE: If you select the `natbib' option in the preamble, far above,
% you should ensure that the \begin{thebibliography} command below is of the 
% conventional LaTeX form
% 
% \begin{thebibliography}{99}
%
% If on the other hand you select the `harvard' option, you should supply
% this command, instead, in the usual Royal Society form
% 
% \begin{thebibliography}
% 
% Note the following `\small' command.  If the `natbib'
% option is used, it is mandatory.  If the `harvard' option is used,
% it is unnecessary, but harmless.

\end{document}